\documentclass[usenatbib]{aa}

\usepackage[switch]{lineno}
\renewcommand\linenumbers{} 

\usepackage[varg]{txfonts}
\usepackage[breaklinks=true]{hyperref} %% to avoid \citeads line fills
\bibpunct{(}{)}{;}{a}{}{,}             %% natbib format for A&A and ApJ

\usepackage[T1]{fontenc}

\usepackage{multirow}
\usepackage{placeins}
\usepackage{url}
\usepackage{array}  % REMOVE IF DONT WORK
\DeclareRobustCommand{\VAN}[3]{#2}
\let\VANthebibliography\thebibliography
\def\thebibliography{\DeclareRobustCommand{\VAN}[3]{##3}\VANthebibliography}

\usepackage{graphicx}	% Including figure files
\usepackage{amsmath}	% Advanced maths commands
\usepackage{amssymb}	% Extra maths symbols

\nolinenumbers %%%%% For ArXiv
\newcommand{\xmm}{{XMM-{\em Newton} }}
\newcommand{\xmmns}{{XMM-{\em Newton}}}
\newcommand{\swift}{{\em Swift }}
\newcommand{\swiftns}{{\em Swift}}

\newcommand{\nustar}{{\em NuSTAR }}

\newcommand{\srcnamelong}{{XMMSL2~J140446.9-251135 }}
\newcommand{\xsrcname}{{XMMSL2~J1404-2511 }}
\newcommand{\xsrcnamens}{{XMMSL2~J1404-2511}}
\newcommand{\msrcnamelong}{{2MASX~14044671-2511433 }}
\newcommand{\msrcname}{{2MASX~1404-25 }}
\newcommand{\msrcnamens}{{2MASX~1404-25}}
\newcommand{\swtd}{{SWIFT~J164449.3+573451 }}
\newcommand{\fluxUnits}{{ergs s$^{-1}$cm$^{-2}$ }}
\newcommand{\fluxUnitsns}{{ergs s$^{-1}$cm$^{-2}$}}
\newcommand{\lumUnits}{{ergs s$^{-1}$ }}
\newcommand{\lumUnitsns}{{ergs s$^{-1}$}}
\newcommand{\msolar}{{$M_{\odot}$ }}
\newcommand{\msolarns}{{$M_{\odot}$}}
\newcommand{\chired}{\chi^{2}_{r}}
\newcommand{\gnh}{$N_{H}=6.7\times10^{20}$ cm$^{-2}$}
\newcommand{\gnhs}{$N_{H}=6.7\times10^{20}$ cm$^{-2}$ }

\newcommand{\eroszerofour}{eRASSt~J045650.3-203750 }
\newcommand{\eroszerofourns}{eRASSt~J045650.3-203750}
\begin{document}
\title{Rapid onset of a
Comptonisation zone in the repeating tidal disruption event \srcnamelong}

% The list of authors, and the short list which is used in the headers.
% If you need two or more lines of authors, add an extra line using \newauthor
\author{
R. D. Saxton\inst{1}\thanks{E-mail: richard.saxton@ext.esa.int} \and
T. Wevers\inst{2,3} \and
S. van Velzen\inst{4} \and
K. Alexander\inst{5} \and
Z. Liu\inst{6,7} \and
A. Mummery\inst{8} \and 
M. Giustini\inst{9} \and
G. Miniutti\inst{9} \and
F. Fuerst\inst{10} \and
J. J. E. Kajava\inst{11} \and 
A. M. Read\inst{12} \and
P. G. Jonker\inst{13} \and 
A. Rau\inst{6} \and
D.-Y. Li\inst{14}
}

% List of institutions
\institute{Telespazio UK for ESA, ESAC, Apartado 78, 28691 Villanueva de la Ca\~{n}ada, Madrid, Spain
\and Astrophysics \& Space Institute, Schmidt Sciences, New York, NY 10011, USA. 
\and Space Telescope Science Institute, 3700 San Martin Drive, Baltimore, MD 21218, USA. 
\and Leiden Observatory, Leiden University, PO Box 9513, 2300 RA Leiden, The Netherlands 
\and Department of Astronomy and Steward Observatory, University of Arizona, 933 North Cherry Avenue, Tucson, AZ 85721-0065, USA
\and Max-Planck-Institut für extraterrestrische Physik, Gießenbachstraße 1, 85748 Garching, Germany 
\and Centre for Astrophysics Research, Department of Physics, Astronomy and Mathematics, University of Hertfordshire, College Lane, Hatfield, AL10 9AB, UK.
\and University of Oxford, Dept. of Physics, Denys Wilkinson building, Keble road, OX1 3RU, Oxford, U.K.
\and Centro de Astrobiolog\'ia (CAB), CSIC-INTA, Camino Bajo del Castillo s/n, ESAC campus, 28692 Villanueva de la Ca\~nada, Madrid, Spain 
\and ESA, ESAC, Apartado 78, 28691 Villanueva de la Ca\~{n}ada, Madrid, Spain
\and SERCO for ESA, ESAC, Apartado 78, 28691 Villanueva de la Ca\~{n}ada, Madrid, Spain
\and Dept. of Physics and Astronomy, University of Leicester, Leicester LE1 7RH, U.K. 
\and Department of Astrophysics/IMAPP, Radboud University, P.O. Box 9010, 6500 GL, Nijmegen, The Netherlands 
\and National Astronomical Observatories, Chinese Academy of Sciences, 20A Datun Road, Chaoyang District, Beĳing, 100101, China 
}

% These dates will be filled out by the publisher
 \date{Received date /Accepted date }

\abstract{
We report here on observations of a tidal disruption event, \xsrcnamens, 
discovered in an XMM-Newton slew, in a quiescent galaxy at $z=0.043$. X-ray monitoring covered the epoch when the accretion disc transitioned from a 
thermal state, with $kT\sim80$~eV, to a harder state dominated by a warm, optically-thick corona. The bulk of the coronal formation took place within 7 days and was coincident with a temporary drop in the emitted radiation by a factor 4. 
%This is consistent with models where power going into the corona causes the disk to collapse and occurred at an accretion rate of 0.05-0.5 Eddington; for higher rates the disk may be radiation-pressure dominated suppressing the formation process. 
After a plateau phase of $\sim 100$ days, the X-ray flux of \xsrcname decayed  by a factor 500 within 230 days.
%, a dropwhich was not due to an increase in line-of-sight absorption. 
We estimate the black hole mass in the galaxy to be
$M_{\rm BH}=(4\pm{2})\times10^{6}$~\msolar and the peak X-ray luminosity 
$L_{\rm X}\sim6\times10^{43}$~\lumUnitsns. The optical/UV light curve is flat over the
timescale of the observations with $L_{\rm opt}\sim 2\times10^{41}$~\lumUnitsns.
We find that TDEs with coronae are more often found in an X-ray sample than in an optically-selected sample. Late-time monitoring of the optical sample is needed to test whether this is an intrinsic property of TDEs or is due to a selection effect. From the fast decay of the X-ray emission we consider that 
the event was likely due to the partial stripping of an evolved star rather than a full stellar disruption,
an idea supported by the detection of two further re-brightening episodes, two and four years after the first event, in the SRG/eROSITA all-sky survey.
}

\titlerunning{Rapid onset of a Comptonisation zone in \xsrcname}
\authorrunning{R. Saxton et al.} 
\maketitle
%%%%%%%%%%%%%%%%%%%%%%%%%%%%%%%%%%%%%%%%%%%%%%%%%%

%%%%%%%%%%%%%%%%% BODY OF PAPER %%%%%%%%%%%%%%%%%%

\section{Introduction}

A tidal disruption event (TDE) occurs when a star passes too close to a supermassive black hole and is disrupted by gravitational tidal forces \citep{Hills75,Rees88,Ulmer99}.
Electromagnetic signals from TDEs have been detected at a wide range of wavelengths including
radio \citep{Alexander20}, infra-red \citep{Jiang16, Li2020}, optical \citep{Arcavi14, Holoien_14ae, Hammerstein23}, UV \citep{Gezari:2003a} and X-rays \citep{KomossaBade:1999a,Esquej:2008a}.
X-ray emission is thought to primarily come from an accretion disc about the SMBH \citep{Rees:1990a,Cannizzo90,Komossa15,MummeryBalbus20,Wen20,Saxton21}
although shocks between streams \citep{Piran:2015br} or between tidal streams and a forming accretion disc \citep{Steinberg22} may contribute at early times, while 
interaction between outflowing material and a pre-existing dense medium has been suggested to contribute on multi-year timescales \citep{Mou21}.

When first observed the X-ray spectrum can often be described purely by thermal disc emission with little
or no hard X-ray emission from a corona \citep[][see \citealt{Mummery23} for a sample]{KomossaBade:1999a,Holoien_14li_6m,Gezari17,Guolo23}.
This is in contrast to active galactic nuclei (AGN) where coronal emission is nearly
ubiquitous \citep{Svoboda17,Laha24}, indicating a fundamental difference in the properties
of newly formed TDE accretion discs from mature persistently-fed discs.
Many TDEs, particularly those discovered in optical surveys, remain in the
thermal state throughout the whole of their monitoring \citep{Guolo23}, sometimes in excess of
a 1000 days \citep{Bright:2018a,Wen23,Wevers23ksf,GuoloMumm25}. In some cases, 
higher cadence monitoring has revealed the formation
of a harder component. AT2018fyk evolved from a thermal
to power-law dominated spectrum in less than a hundred days \citep{Wevers21} 
%and displayed a subsequent softening contemporary with a dramatic late-time fading of the source.
while AT2021ehb gradually formed a dominant hard spectral component over 170 days which softened significantly within 3 days while the X-ray flux faded by an order of magnitude \citep{Yao22}.
Radio observations of these sources showed that the hard X-ray component is not related to a relativistic jet as was found to be the case for \swtd \citep{Burrows11,Bloom11,Zauderer:2013a} 
\citep[but see][]{Christy24}. 
The interpretation is that 
hard X-rays are generated by the up-scattering of disc photons by a corona \citep[e.g.][]{Wevers:2019b}.
%which may be created by the amplification of initial magnetic fields frozen into the stellar debris
%by the MRI mechanism \citep{BalbusHawley91}. 
It is unclear why this component only appears to develop in a minority of TDEs,
although the time needed to magnetically build the corona 
may be a factor \citep{Yao22}.

These events give a view of the dynamic evolution of accretion structure
around SMBH, which is otherwise only accessible in the rare
changing-look AGN \citep[CLAGN;][]{Ricci22}. 
State changes are particularly prevalent in partial TDEs \citep[pTDE;][]{Guillochon13}, some of which show multiple peaks in their light curves indicating periodic 
stripping of material from an evolved star \citep{Wevers23fyk,Liu24}.
This can be understood in terms of each stripping providing a relatively low mass of material 
to the system, resulting in a rapid change of the accretion rate
as material is consumed. Changes in accretion rate are believed to influence
state changes seen in stellar mass black hole discs \citep{Esin97}.

In this paper we describe a TDE discovered in the \xmm slew survey,
\xsrcnamens, whose X-ray spectrum demonstrates a rapid build up of the
corona shortly after detection. 

In Section 2 we discuss the discovery of this TDE and the source identification;
in Sections 3, 4 \& 5 we present UV, optical, radio and X-ray follow-up 
observations and in Section 6 we discuss the source
characteristics within the TDE model. The paper is summarised in Section 7.

A $\Lambda$CDM cosmology with ($\Omega_{M},\Omega_{\Lambda}$) = (0.27,0.73)
and  $H_{0}$=70 km$^{-1}$s$^{-1}$ Mpc$^{-1}$ has been assumed throughout.

\section{Discovery of \xsrcname}

\begin{figure}
    \resizebox{\hsize}{!}{\includegraphics{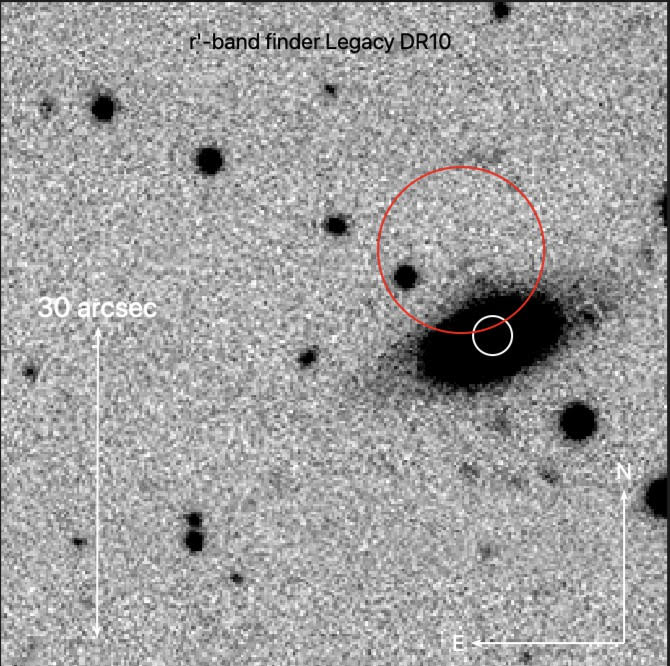}}
    \caption{Finder chart of \xsrcname based on an r' band image from the
    Legacy DR10 survey performed on DECam on the Blanco CT10 telescope.
    %\footnote{https://www.legacysurvey.org/dr10/description/}. 
    Red circle is the \xmm slew error circle of 8 arcseconds radius and the white circle is the Swift-UVOT enhanced position with an error of 1.9 arcseconds.}
    \label{fig:dss}
\end{figure}

A new source of X-rays, \srcnamelong (hereafter \xsrcnamens), was discovered by \xmm \citep{Jansen:2001a} while it slewed between targets on Feb 15 2018, with a count rate of $4.4\pm{1.1}$ c/s in the EPIC-pn camera \citep{Atel11394}.
The X-rays were located to a position RA: 14 04 46.68 DEC: -25 11 43.3 ($\pm1.9$ arcsecs; 90\% confidence) by follow-up observations with the Neil Gehrels Swift observatory \citep[henceforth \swiftns;][]{Gehrels:2004a} and shown to be coincident with the galaxy \msrcnamelong (Fig.~\ref{fig:dss}).

The EPIC-pn slew spectrum was soft and could be fit with a black-body of $92^{+30}_{-22}$\,eV and 0.2--2\,keV flux $5.6\times10^{-12}$\,\fluxUnits absorbed by the Galactic column
of \gnh \citep{Willingale13}. The ROSAT All Sky Survey (RASS) 2-sigma upper limit from 1990 was 0.033 c/s equating to a flux of $<5\times10^{-13}$\fluxUnits using the same spectral model, a factor $>10$ lower than the slew detection. 

\section{Optical and UV follow-up}
\label{Sec:optspec}

\begin{figure}
        \resizebox{\hsize}{!}{\includegraphics{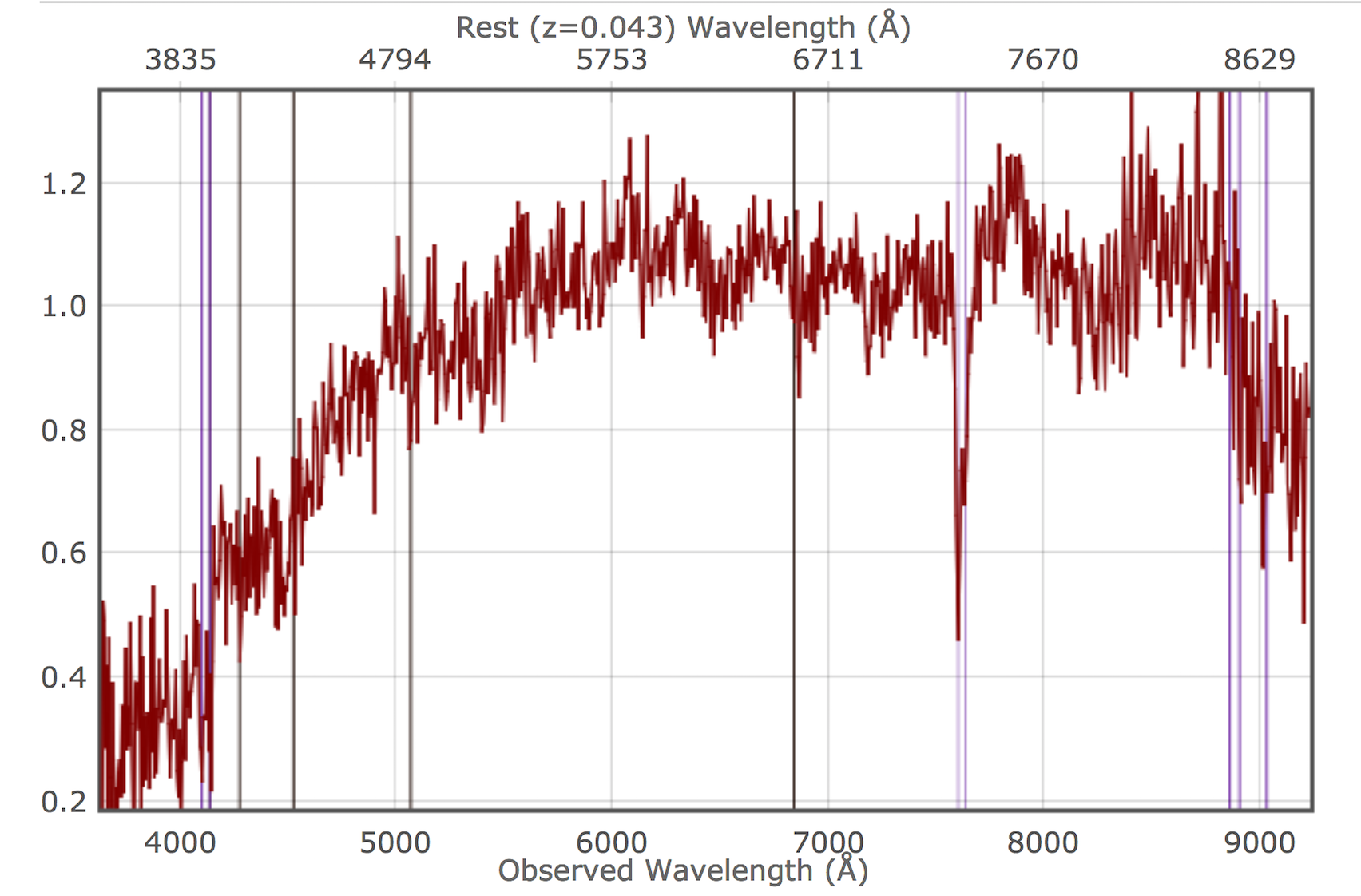}}
    \caption{Optical spectrum of \msrcname taken with the New Technology Telescope (NTT) on 2018-03-08
    as part of the ePESSTO program.}
    \label{fig:ePESSTO}
\end{figure}

An optical spectrum was taken by the NTT within the ePESSTO program \citep{Smartt15} on 2018-03-08 revealing a quiescent galaxy with no emission lines at redshift $z=0.043\pm{0.001}$ (\citealt{ATel11395}; Fig.~\ref{fig:ePESSTO}).

\begin{figure}
        \resizebox{\hsize}{!}{\includegraphics{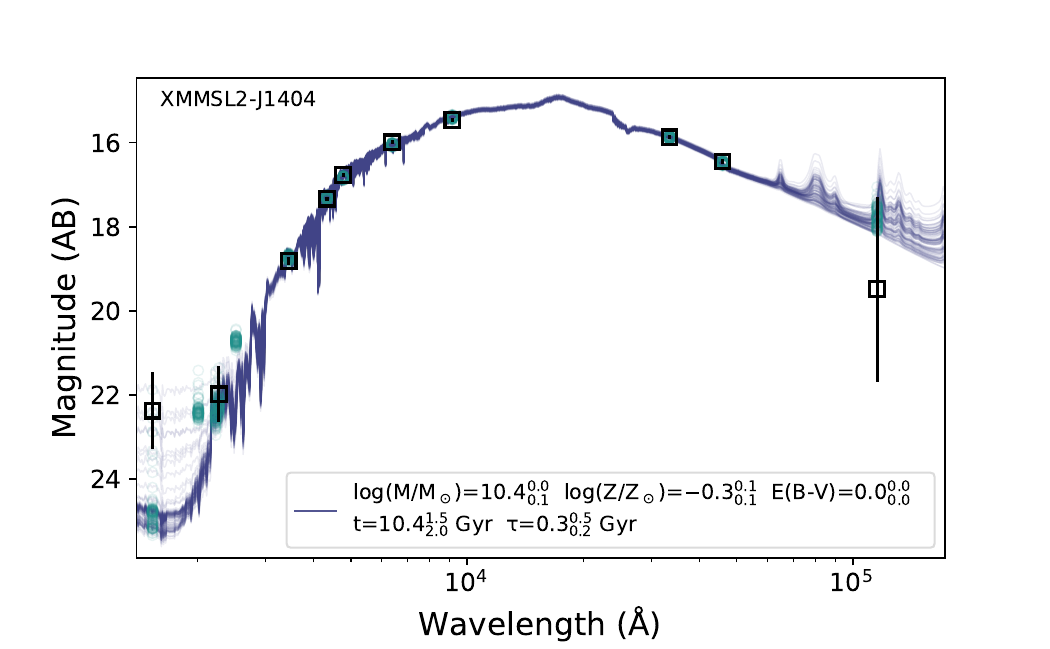}}
    \caption{
    Host galaxy photometry based on GALEX \citep{Martin05}, DECaLS \citep{Dey19}, and WISE \citep{Wright10,Mainzer11}.  We also include the post-peak aperture photometry in the UVOT B- and U-band. We show samples from the posterior distribution of population synthesis galaxy models \citep{Conroy09,Johnson17}, following the method outlined in \citet{vanVelzen21}. The predicted magnitudes (including UVOT W1, M2, and W2) are shown with open circles.}  
    \label{fig:magellan}
\end{figure}

A further optical spectrum was taken with the MagE instrument on the Magellan
telescope with a 0.7 arcsecond slit on 2020-02-15 (\citealt{wevers20}; Fig.~\ref{fig:magellan}).
Blaze corrections were performed for each individual échelle order by fitting a low order spline function to regions devoid of absorption and emission lines, after which the spectrum was stitched together using an inverse variance weighting scheme for overlapping wavelength ranges between orders. Following the procedures outlined in \citet{wevers2017}, the Penalized Pixel fitting routine \citep{cappellari2017} was used in combination with the Elodie stellar template library \citep{prugniel2007} to measure the velocity dispersion of this continuum normalized spectrum, correcting for the instrumental broadening of $\sigma_{\rm inst}$ = 22 km s$^{-1}$. A velocity dispersion of $93\pm{1}$ km/s was measured. Using the $M-\sigma$ relation of \cite{Gultekin09} a black hole mass of log$_{10}(M_{BH})=6.71\pm{0.4}$\msolar is obtained 
with the error dominated by the systematic error of the method\footnote{Using an alternative scaling relation \citep{Greene20} gives a consistent mass of log$_{10}(M_{BH})=6.80\pm{0.55}$\msolar.}.

\begin{figure}
     \resizebox{\hsize}{!}{\includegraphics{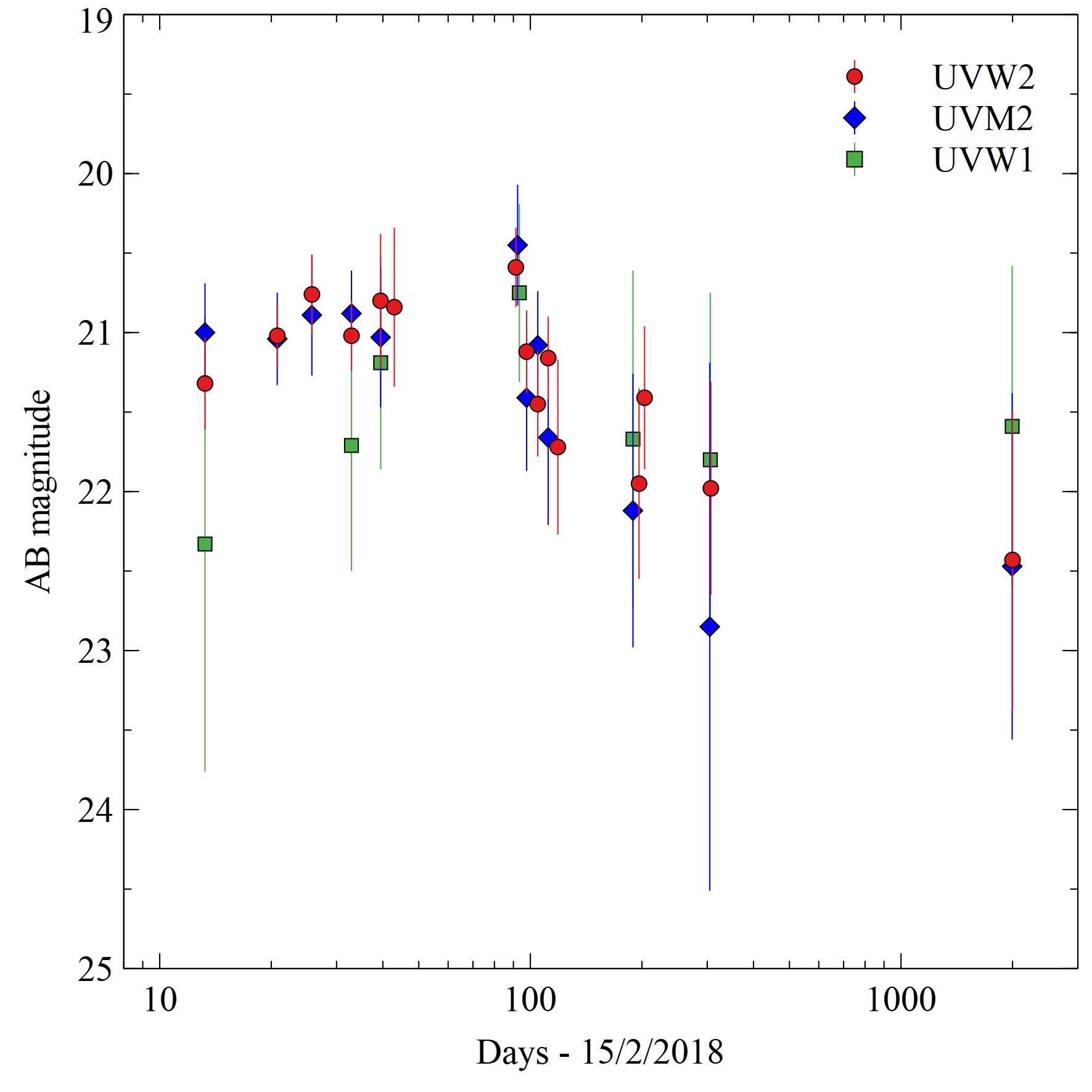}}
     \resizebox{\hsize}{!}{\includegraphics{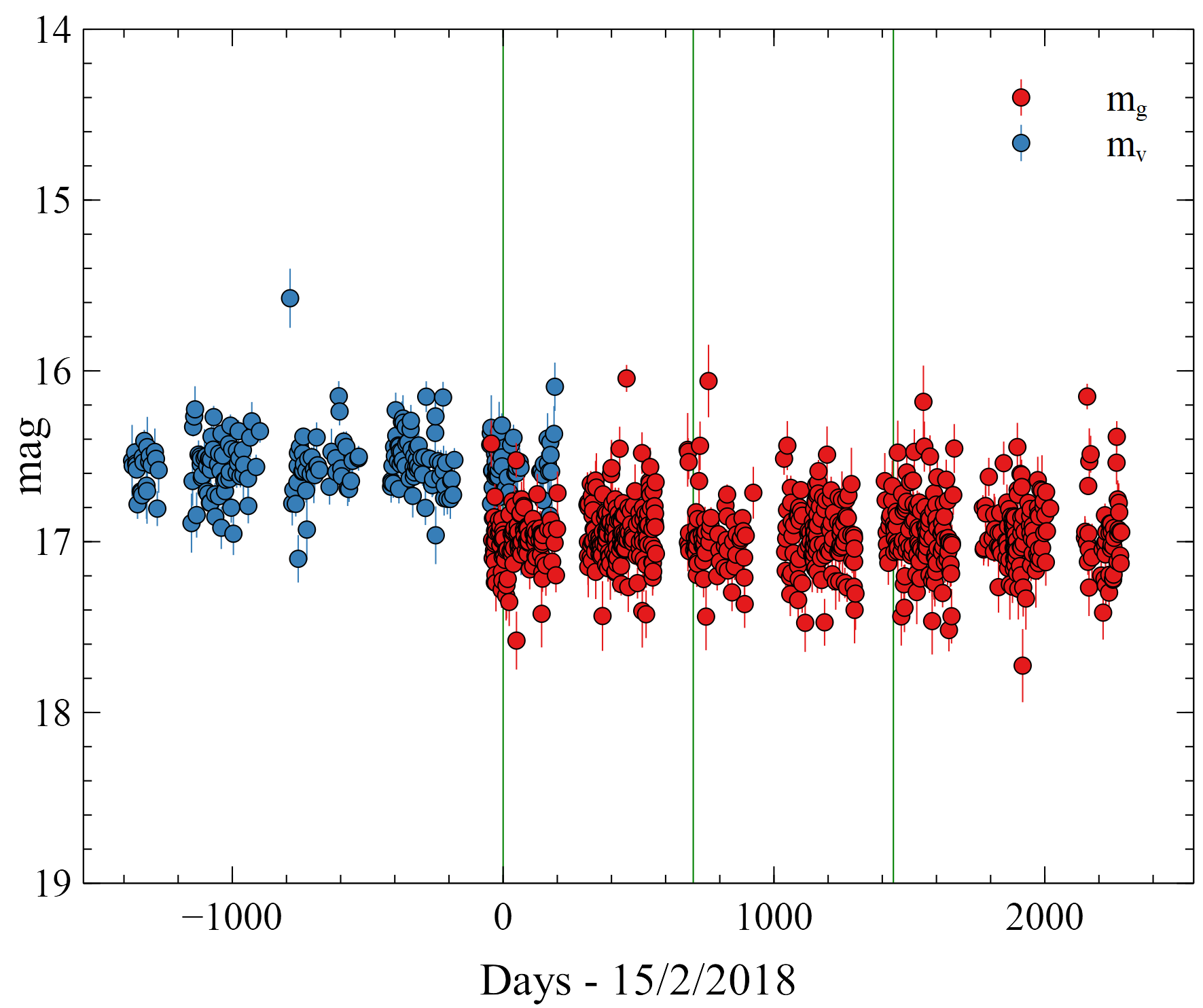}}
    
    \caption{Light curve of \xsrcname in the Swift-UV filters (upper) and
    in V and G filters from the ASAS-SN
    Sky Patrol v2.0 project \citep{Hart23,Shappee14} (lower) where
    the vertical green lines indicate the dates of 
    the three flares seen in X-ray data. Magnitudes represent the 
    host galaxy light, which dominates in the optical filters, plus
    light from the TDE which dominates in the UV.}
    \label{fig:optuvlc}
    \label{fig:asassn}
\end{figure}

Photometric observations were made with the Swift ultraviolet telescope (UVOT; Tab.~\ref{tab:uvobs}) and with the XMM-Newton optical monitor (XMM-OM). In both
telescopes the B, U, UVW1, UVM2 and UVW2
filters were used and UVOT also observed with the V filter (Fig.~\ref{fig:optuvlc}). The galaxy was faint being detected in all of the UVOT filters but not in the XMM-OM camera in the UVW2 or UVM2 filters. 
The UV temperature, obtained by fitting all UVOT data simultaneously to a power-law decay and a
single black-body is $T = 10^{4.4\pm{0.1}}$ K, which is at the high end of the distribution
of temperatures seen in a sample of 30 spectroscopically-classified ZTF TDEs 
\citep{Hammerstein23}. 
%I see some evidence for a turnover in the W2 band, so a thermal spectrum is likely preferred over a single power-law (SVV).
The black-body integrated optical/UV luminosity of $L_{\rm bb}\sim2\times10^{41}$\lumUnits evolves slowly and if the optical emission did have a peak then that peak would have occurred well before the first X-ray observation (Fig.~\ref{fig:optuvlc}).

The GALEX catalogue \citep{bianchi17} contains a source (GALEX~J140446.5-251140) at an offset of 4 arcseconds from \msrcnamens, detected in the NUV
filter with magnitude $21.81\pm{0.65}$ on 2007-05-12 but undetected in the FUV filter. 

Historical optical observations from the All-Sky Automated Survey for SuperNovae \citep[ASAS-SN;][]{Shappee14} are flat from 2014 until 2022, although we note that the light curve has a gap between 2017-08-19 and 2017-12-17 (Fig.~\ref{fig:optuvlc}) . 
No Gaia science alert \footnote{http://gsaweb.ast.cam.ac.uk/alerts/alertsindex}
was triggered for \xsrcname also indicating that the source has not brightened by more than 1 magnitude in the Gaia G band since the alerts started in 2016 \citep{Hodgkin21}.

% UV/opt light curve plot

\begin{center}
\begin{table}
{\small 
\caption{Swift-UVOT observations of \xsrcname}
\label{tab:uvobs}
\hfill{}
\begin{tabular}{l c c c c}
\hline\hline
Date & U & UVW1 & UVM2 & UVW2 \\
 \hline
% Update value - from SVV, 21/05/24
13 & $20.93\pm{0.76}$ & $21.62\pm{0.69}$ & $20.80\pm{0.24}$ & $21.18\pm{0.23}$\\
21 & $21.01\pm{0.78}$ & $21.97\pm{0.87}$ & $20.84\pm{0.22}$ & $20.91\pm{0.18}$\\
26 & $21.73\pm{2.09}$ & $22.85\pm{2.80}$ & $20.72\pm{0.31}$ & $20.68\pm{0.22}$\\
33 & - & $21.25\pm{0.48}$ & $20.71\pm{0.21}$ & $20.92\pm{0.19}$\\
39 & - & $20.89\pm{0.49}$ & $20.83\pm{0.35}$ & $20.71\pm{0.38}$\\
43 & $19.95\pm{0.73}$ & $21.53\pm{1.53}$ & $21.74\pm{1.27}$ & $20.75\pm{0.46}$\\
51 & - & $21.07\pm{0.84}$ & $22.08\pm{1.37}$ & $22.55\pm{1.59}$\\
92 & $20.79\pm{1.09}$ & $20.54\pm{0.45}$ & $20.33\pm{0.34}$ & $20.52\pm{0.23}$\\
98 & $21.74\pm{1.75}$ & $21.55\pm{0.70}$ & $21.13\pm{0.33}$ & $21.01\pm{0.22}$\\
105 & - & $21.62\pm{0.76}$ & $20.87\pm{0.27}$ & $21.29\pm{0.27}$\\
112 & $22.44\pm{3.19}$ & - & $21.33\pm{0.38}$ & $21.04\pm{0.21}$\\
119 & $21.09\pm{1.40}$ & $21.47\pm{0.97}$ & $21.22\pm{0.66}$ & $21.52\pm{0.44}$\\
189 & - & $21.23\pm{0.68}$ & $21.64\pm{0.52}$ & $22.64\pm{0.97}$\\
196 & $21.06\pm{1.14}$ & $21.88\pm{1.16}$ & $22.34\pm{0.99}$ & $21.72\pm{0.45}$\\
203 & $20.49\pm{0.75}$ & - & - & $21.26\pm{0.38}$\\
208 & - & - & $21.66\pm{1.11}$ & $21.45\pm{0.85}$\\
306 & - & $21.31\pm{0.64}$ & $22.06\pm{0.74}$ & $21.74\pm{0.51}$\\
1989 & - & $21.46\pm{0.81}$ & $22.54\pm{1.02}$ & $22.34\pm{0.87}$\\
1995 & $20.72\pm{1.05}$ & $21.17\pm{0.66}$ & $21.86\pm{0.57}$ & $22.06\pm{0.69}$\\
1999 & - & - & $22.26\pm{0.68}$ & $22.08\pm{0.65}$\\
2301  & - & - & - & $21.87\pm{0.36}$ \\
2309 & - & - & - & $22.34\pm{0.61}$ \\
2319 & - & - & - & $21.23\pm{1.11}$ \\
\hline
\end{tabular}
\hfill{}
\\
\tablefoot{
Swift-UVOT host-subtracted, Galactic-extinction corrected AB magnitudes.
Date specifies the number of days since 15/02/2018. }
}
\\\end{table}
\end{center}

\section{Radio observations}
\label{sec:radio}

\xsrcname was observed with NSF's Karl G Jansky Very Large Array (VLA) at a mean frequency of 6 GHz on 2018 March 23 and again on 2018 September 8 under program 18A-453 (PI: Alexander). In the first observation the VLA was in its highest resolution A configuration, while in the second it was in its lowest resolution D configuration. Both observations used 3C286 as the flux calibrator and J1409-2657 as the phase calibrator and the data were reduced in CASA following standard procedures. 

\xsrcname was not detected in either observation, with $3\sigma$ limits of 14 $\mu$Jy and 35 $\mu$Jy respectively.

\section{X-ray observations and long-term light curve}

After discovery, a monitoring program was immediately initiated with \swift with a cadence of one observation per week. Despite being discovered in an \xmm slew the source had formally left the
\xmm visibility window at the time and could only be observed by \xmmns, in a pointed observation, five months later (OBSID=0804860201).
A second pointed observation was made after a further six months (OBSID=0804860301).
The full list of X-ray observations is given in Tab.~\ref{tab:xobs}.

\swift spectral products were extracted using the on-line XRT data products tool available from the UK Swift Data centre \citep{Evans09}. As another source lies at $\sim$~1 arcminute from \xsrcname we entered the source coordinates for the fainter observations to avoid problems with the centroiding routine.
A brief analysis of the secondary source spectrum is presented in Appendix A.

Data from the first \xmm observation were reduced using SAS 18.0
\citep{Gabriel04}. Source products were extracted from a circle of radius 27 arcseconds 
about the source and a local background was taken from the same CCD. The background was low throughout the observation and flare screening was not applied.
We restricted our analysis to the EPIC-pn camera.

\begin{figure}
    \resizebox{\hsize}{!}{\includegraphics{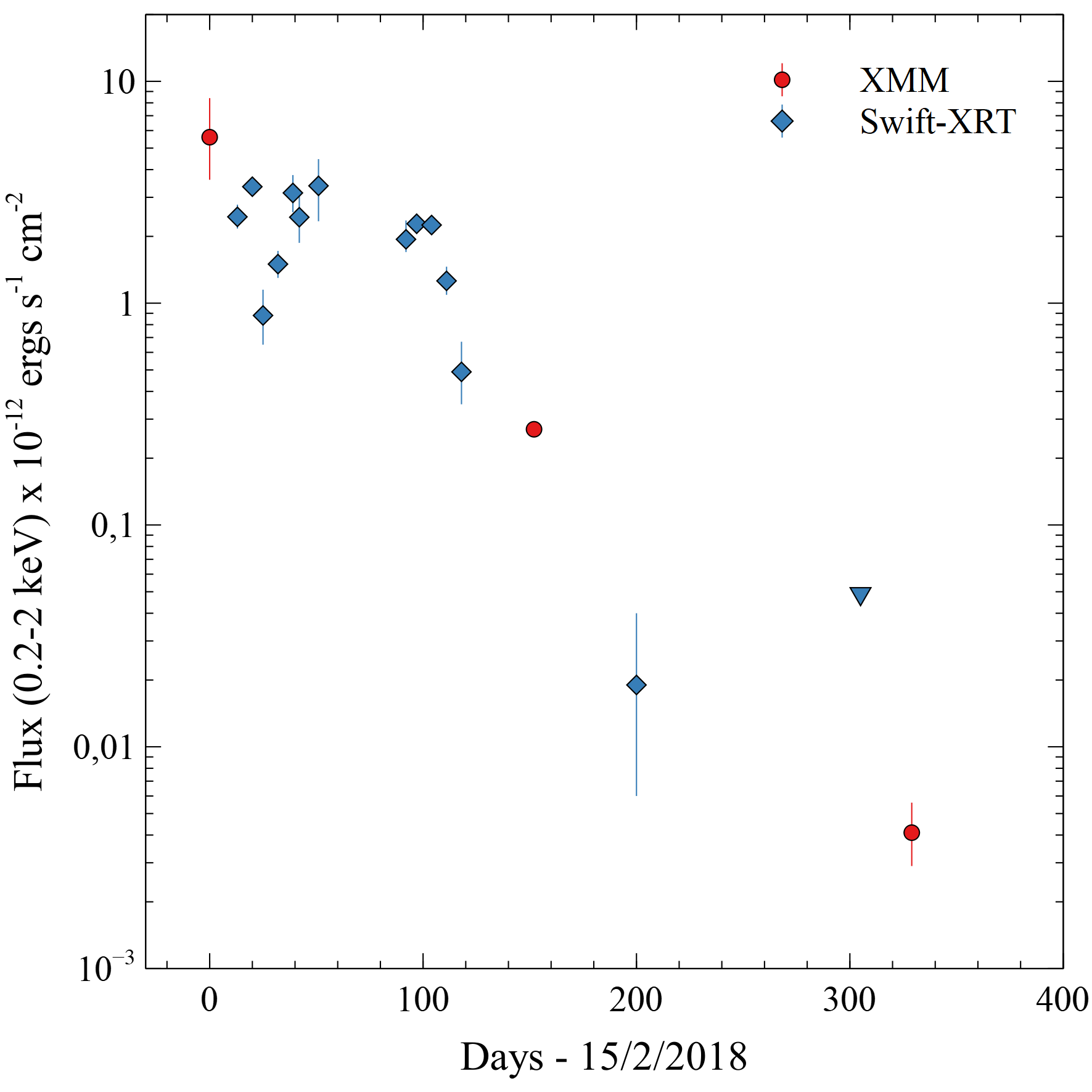}}
 
    \caption{Observed soft X-ray (0.2-2 keV) flux light curve of \xsrcname during 2018 and 2019. XMM-Newton points are shown as red
circles and Swift-XRT points as blue diamonds or a blue downward triangle for the upper limit.
Fluxes have been calculated using the spectral model {\em COMPBB} (see Tab.~\ref{tab:specfits}).
}
    \label{fig:Xlc_longterm}
\end{figure}

The long-term 0.2-2 keV observed flux light curve is given in Fig.~\ref{fig:Xlc_longterm} 
and can be characterised by a flat flux level for the first 100 days followed by a decay of $t^{-5.2\pm{0.2}}$. 
A similar 100-day initial flat period followed by a constant 
decay was seen in XMMSL1 J1446+68  \citep{Saxton:2019a}, although the rate of decline is considerably steeper 
in \xsrcnamens, of the order of a factor 500 in 230 days.
This decay is notably faster than observed in most previous events \citep{auchettl17} including ASSASN-14li \citep{Brown17,Wen20}.

\begin{center}
\begin{table}
{\small 
\caption{X-ray observation log of \xsrcname}
\label{tab:xobs}
\hfill{}
\begin{tabular}{l l c l l }
\hline\hline
Mission & Key$^{a}$ & Date & Expos & Count rate$^{b}$ \\
 &  & & (s) &  \\
 \hline
ROSAT & R1 & 1991-01-07  & 320 & $<0.033$ \\  
XMM slew & XS1 & 2018-02-15 & 3.4 & $4.4\pm{1.1}$ \\ 
Swift-XRT & S1 & 2018-02-28 & 2856 & $ 0.084\pm 0.006$ \\
Swift-XRT & S2 & 2018-03-07 & 3157 & $0.124 \pm 0.007$ \\
Swift-XRT & S3 & 2018-03-12 & 1394 & $0.033 \pm 0.007$ \\
Swift-XRT & S4 & 2018-03-19 & 3064 & $0.066 \pm 0.006$ \\
Swift-XRT & S5 & 2018-03-26 & 950 & $0.16 \pm 0.02$ \\
Swift-XRT & S6 & 2018-03-29 & 436 & $0.10 \pm 0.02$ \\
Swift-XRT & S7 & 2018-04-07 & 682 & $0.11 \pm 0.02$ \\
Swift-XRT & S8 & 2018-05-18 & 1148 & $0.084 \pm 0.010$ \\
Swift-XRT & S9 & 2018-05-23 & 2788 & $0.096 \pm 0.007$ \\
Swift-XRT & S10 & 2018-05-30 & 2883 & $0.104 \pm 0.007$ \\
Swift-XRT & S11 & 2018-06-06 & 2928 & $0.060 \pm 0.005$ \\
Swift-XRT & S12 & 2018-06-13 & 1427 & $0.023 \pm 0.005$ \\
XMM pnt & X1 & 2018-07-17 & 32900 & $0.217 \pm 0.003$ \\
Swift-XRT & S13-16$^{c}$  & 2018-09-02 & 8064 & $6.5\pm{3.3}\times10^{-4}$ \\
Swift-XRT & S17 & 2018-12-17 & 2788 & $<0.0020$ \\
XMM pnt & X2 & 2019-01-10 & 29874 & $3.5\pm{0.7}\times10^{-3}$ \\
NuSTAR    & N1 & 2019-01-10 & 68557 & $<0.0024^{d}$ \\
XMM slew & XS2 & 2020-07-17 & 2.1 & $<1.72$ \\
eROSITA & eRASS1 & 2020-01-26 & 321 & $0.23\pm{0.03}^{e}$\\   
eROSITA & eRASS2 & 2020-07-28 & 320 & $0.09\pm{0.02}$ \\
eROSITA & eRASS3 & 2021-01-20 & 222 & $<0.02$ \\
eROSITA & eRASS4 & 2021-07-28 & 219 & $<0.04$ \\
eROSITA & eRASS5 & 2022-01-26 & 232 & $0.25\pm{0.03}$ \\
Swift-XRT & S18-21$^{f}$ & 2023-08-02 & 9991 & $<0.00043$ \\
Swift-XRT & S22-23$^{g}$ & 2024-06-07 & 2420 & $<0.0012$ \\
\hline
\end{tabular}
\hfill{}
\\
$^{a}$ R=ROSAT survey, XS=XMM-Newton slew, S=Swift-XRT, X=XMM-Newton pointed, N=NuSTAR, eRASS=eROSITA survey. \\
$^{b}$ Detector count rate (counts/s) in the 0.2--2 keV energy band. Upper limits are 2-$\sigma$. \\
$^{c}$ Summation of 4 observations made on 2018-08-23, 2018-08-30, 2018-09-06 and 2018-09-11. \\
$^{d}$ Upper limit to the count rate from the combined FPA and FPB \nustar detectors in the energy band 3--20 keV.\\
$^{e}$ All eROSITA count rates and upper limits are background subtracted but have not
been corrected for vignetting or encircled energy effects. \\
$^{f}$ Summation of 4 observations made on 2023-07-28, 2023-08-02, 2023-08-03, 2023-08-07. \\
$^{g}$ Summation of 2 observations made on 2024-06-04 and 2024-06-11. \\
}
\end{table}
\end{center}

\subsection{X-ray spectral analysis}
\label{sec:specfits}
\begin{figure}
    \resizebox{\hsize}{!}{\includegraphics{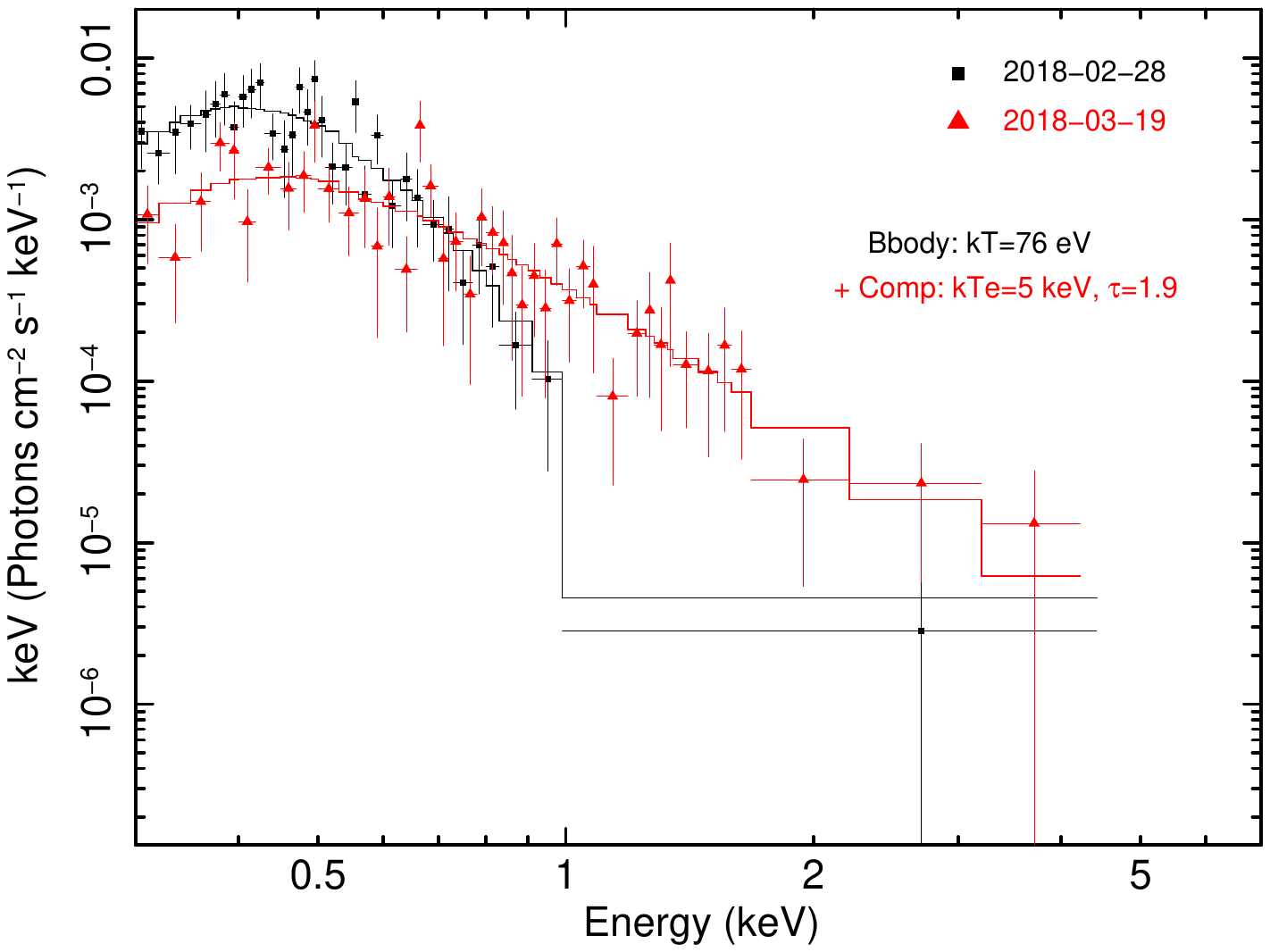}}
    
    \caption{Swift-XRT spectra of \xsrcname from 2018-02-28 fitted with a black-body (black) and 2019-03-19 fitted with a COMPBB model (red).}
    \label{fig:RapidSpecChange}
\end{figure}

\begin{figure}	
       \resizebox{\hsize}{!}{\includegraphics{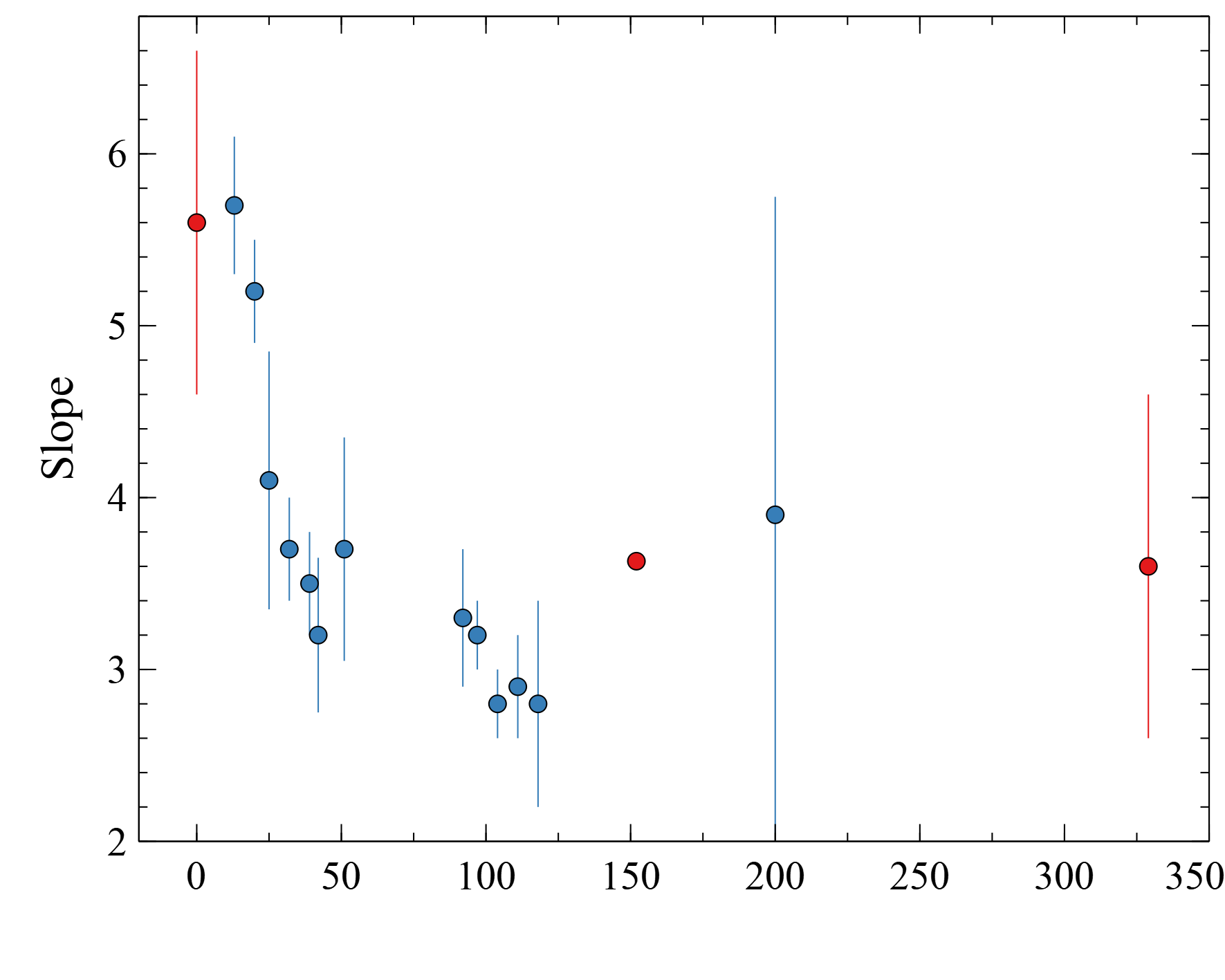}}
        \resizebox{\hsize}{!}{\includegraphics{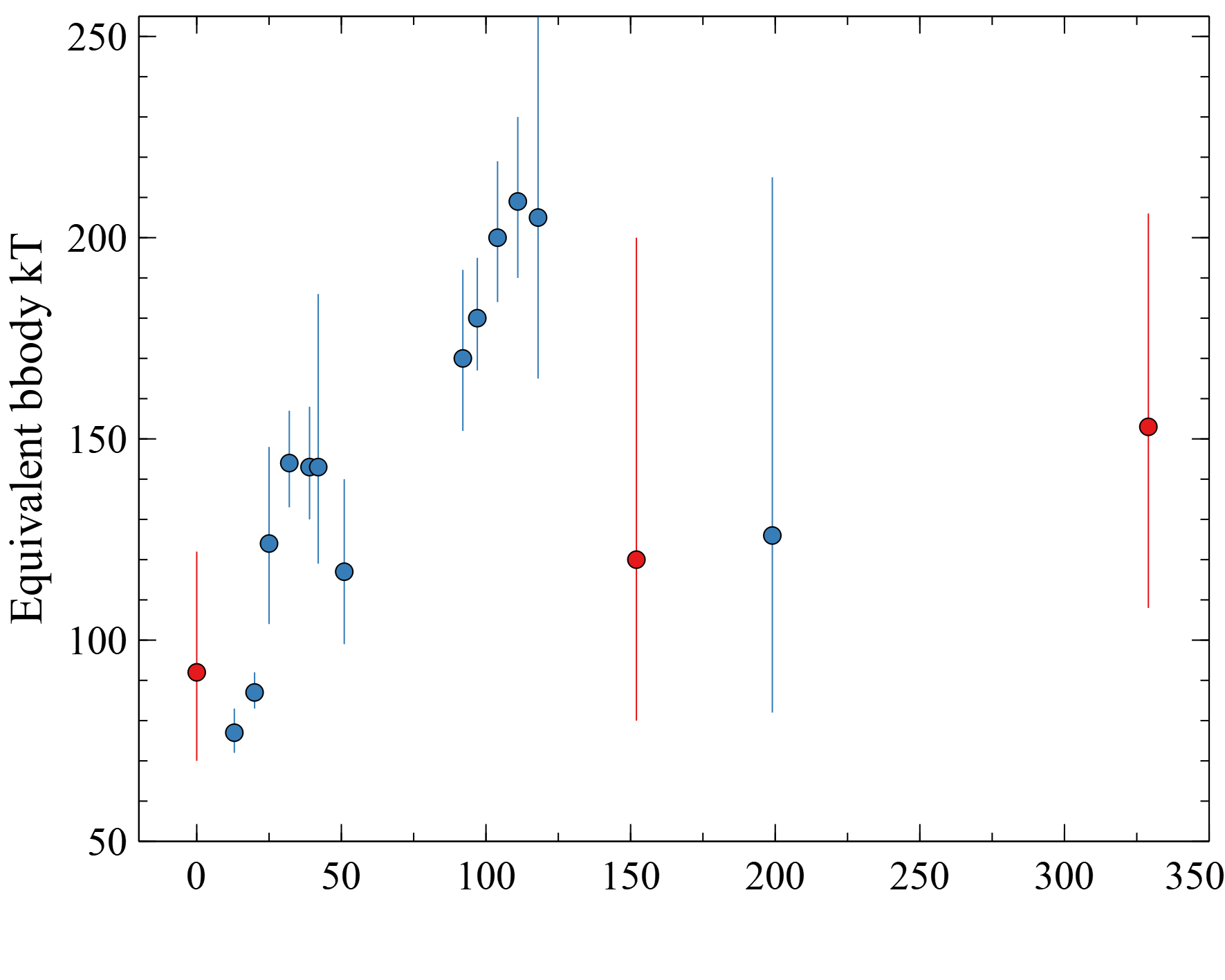}}
        \resizebox{\hsize}{!}{\includegraphics{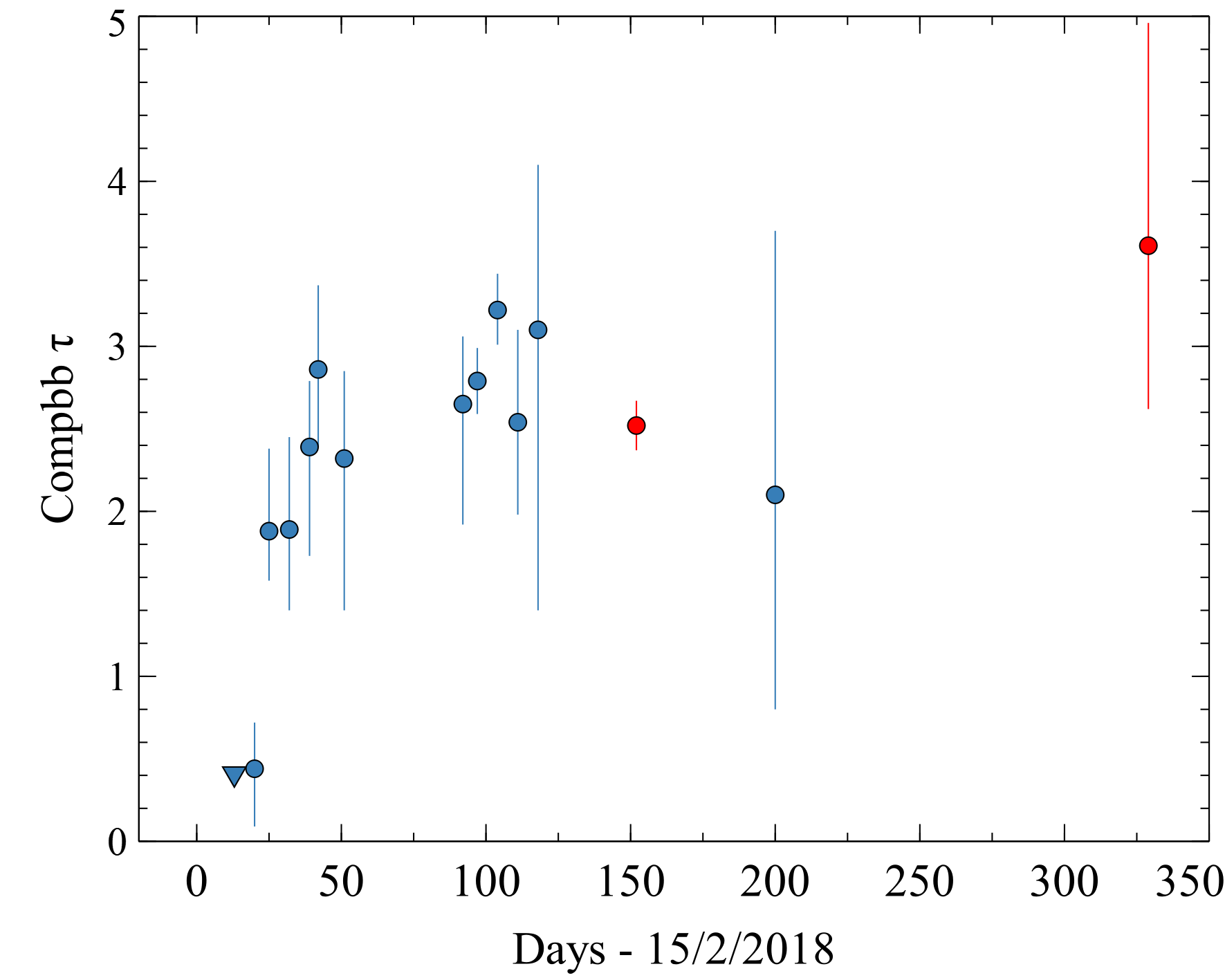}}
	\caption{Upper: Slope of a power-law model absorbed by the Galactic column fitted to the X-ray spectra of \xsrcnamens. Red points are from XMM-Newton, blue points from Swift-XRT.
    Middle: Temperature (eV) of a simple model of a black-body absorbed by the Galactic column
of \gnh fitted to the X-ray spectra of \xsrcnamens. Note that this does not represent the physical temperature of the disc but illustrates how the spectrum hardens with time. 
Lower: Optical depth of a 5 keV electron population derived from fits of a {\em COMPBB} model, 
absorbed by the Galactic column, to X-ray spectra of \xsrcname from 2018 and 2019.}
    \label{fig:SimpleSpecFits}
    \label{fig:CompFits}
\end{figure}

The early Swift-XRT observations of 2018-02-28 and 2018-03-07 can be well fit
with a single black-body model of temperature kT$\sim80$ eV, absorbed by the Galactic
column of \gnh. TDEs at the peak of their emission commonly
show a similar thermal spectrum \citep{KomossaBade:1999a,Esquej:2008a,Brown17,Guolo23}. 
However, in subsequent observations the spectrum of \xsrcname hardened and became more complex. As an illustration, in
Fig.~\ref{fig:RapidSpecChange} we show the evolution of the Swift-XRT spectrum between the
observations of 2018-02-28 and 2018-03-19, where the hardening is very evident. 
To visualise the evolution of the spectrum, we parameterise the hardness of
the X-ray spectrum over the full range of observations by the slope returned in a simple
fit of a power-law, absorbed by the Galactic column,  
to the 0.3-10 keV spectra from XMM-Newton and Swift-XRT (Fig.~\ref{fig:SimpleSpecFits}).
This model does not give a particularly good fit but clearly illustrates the evolution as
the spectral index drops from $\Gamma\sim5.5$, twenty days after discovery, to a minimum of $\Gamma<3$ on days 100--120.
Alternatively, we can parameterise the spectral hardness by fitting a single absorbed black-body to the same spectra 
(Fig.~\ref{fig:SimpleSpecFits}). The temperature of the black-body apparently increases from $\sim80$ eV until it peaks at $\sim200$ eV between days 100--120, 
in parallel with the changes seen in the power-law slope. We will see later that this does not 
represent the real temperature of the disc emission and point out here that care must be taken when 
interpreting measured X-ray temperatures in terms of real-world parameters, especially in low s/n spectra.
In both parameterisations, the three observations taken between days 150 and 329 (X1,S13-16,X2) show tentative evidence for a softening of the spectrum; this is most obvious in the simple power-law fit where the slope steepens from 2.8 (in S12) to 3.6 (in X1).

A visual inspection of Fig.~\ref{fig:RapidSpecChange} shows that the flux
below 0.7 keV declined between 2018-02-28 and 2018-03-19, while the flux
above 0.7 keV increased. This immediately suggests that the soft photons have been reprocessed into harder photons. 

We used the high-statistic XMM-Newton observation of 2018-07-17 (day 150), using the EPIC-pn data,
grouped to have a minimum of 25 counts in each spectral bin, to investigate
the spectral model. 
Spectral fits were performed with {\em XSPEC v12.13.1} \citep{xspecref}. All fits include absorption by the 
Galactic column (\gnh) with Wilm abundances \citep{WilmABund}.

First we tried single models, finding a poor fit 
with an absorbed black-body ($\chired=546/146$), bremsstrahlung ($\chired=251/146$)
or a multi-colour disc (DISKBB model; $\chired=392/146$).
A power-law proved a better fit ($\chired=166/146$, with $\Gamma=3.63\pm{0.05}$).
A good fit ($\chired=147/144$) was given by a black-body
Comptonised by a warm electron population \citep[COMPBB;][]{Nishimura:1986a} with a rest-frame black-body
temperature of $kT=48^{+16}_{-10}$ eV, electron
temperature $kT_{\rm e}=2.7^{+6.2}_{-2.4}$ keV, optical depth $\tau=3.8^{+1.3}_{-2.3}$ and
normalisation of $5.3\times10^{4}$.
This model treats soft X-ray photons Compton-scattered by a plane-parallel plasma. 
Switching to a different comptonisation model \citep[THCOMP;][]{thcomp20} gave a similar quality fit with comparable temperatures and optical depth.

As a further check we added an extra component of intrinsic cold absorption ({\em ztbabs}), at the redshift
of the source, finding no improvement in the fit ($\chired=147/143$; 
$N_{H}<1.5\times10^{20}$ $cm^{-2}$). A warm absorber 
\citep[{\em absori}; ][]{Done_absori, Mag_absori} likewise gave no significant
improvement ($\chired=147/142$).

We then applied the {\em COMPBB} model to the lower-statistic \swiftns-XRT observations, fixing the electron temperature
to kT$_{e}$=5 keV. We choose this temperature as being close to the centre of the allowed
range from the \xmm observation and note that it is hotter than the 0.1--1.0 keV temperature
derived for the soft excess in a sample of AGN \citep{Petrucci18} but well below the 
temperature of the hot corona \citep[e.g.][]{Akylas21}. Results are presented in table~\ref{tab:specfits}. The evolution of the optical depth
of the electron corona is shown in Fig.~\ref{fig:CompFits} and jumps from $\tau<0.4$ to an optical depth
$\tau>1.6$ in 12 days between 2018-02-28 and 2018-03-12 or from $\tau=0.4\pm{0.3}$
to $\tau=1.9^{+0.5}_{-0.3}$ in 5 days from 2018-03-07 to 2018-03-12.
Overall the corona went from being undetectable to being fully formed, with a depth
$\tau=2.9\pm{0.5}$, in 29 days between 2018-02-28 and 2018-03-29.

Over the set of observations, the observed temperature of the disc emission dropped from 
$\sim80$ eV in the early
Swift observations to $kT=50\pm{10}$ eV (with kT$_{e}$ fixed at 5 keV) in the XMM-Newton observation of 2018-07-17 (Tab.~\ref{tab:specfits}).
%and for the observations where it can be constrained follows the L$\propto T^{4}$ law (Fig.~\ref{fig:LvT4}) 
%suggesting constant-area thermal emission likely associated with the inner regions of an accretion disc.

As an alternative we investigated whether a high-temperature (fixed to 70 keV), low optical depth,
Comptonisation model could also fit the X1 spectrum. Using the {\em COMPBB} model the fit is bad
(267/145), but changing to the {\em COMPTT} model \citep{Titarchuk:1994a} does allow a
reasonable high-temperature (frozen to 70 keV), $\tau=0.015\pm{0.003}$ fit to the EPIC-pn data ($\chired=151/144$). As noted elsewhere \citep[e.g.][]{Akylas21,Tamborra18}, extracting physical parameters
for the electron population in the corona is very model dependent unless a high-energy break is available to constrain the electron temperature.

% Plot of spectral fit to X1 observation
%\begin{figure}
	
%    \resizebox{\hsize}{!}{\includegraphics{xmm1_compbb.pdf}}
   
%	\caption{Spectral fit of a redshifted, comptonised black-body ($N_{\rm H}$*ZASHIFT*COMPBB) to the XMM-Newton spectrum of \xsrcname taken on 2018-07-17. Solid black line traces the model with kT=48eV, kTe=2.7keV and $\tau=3.8$. }
%    \label{fig:xmm1_specfit}
%\end{figure}

% Table of spectral fits to a comptonised black-body
\begin{table*}%[ht]
\begin{center}
{\small
\caption{Fits of a Comptonised black-body to \xmmns, \swiftns-XRT and SRG/eROSITA observations of \msrcnamens}
\label{tab:specfits}  

% is used to refer this table in the text
\begin{tabular}{llllll}
\hline\hline                 % inserts double horizontal lines
Date & Key & $kT^{a}$ & $\tau^{b}$ & Flux (0.2--2\,keV) & C/dof \\
         & & eV &  & $10^{-12}$\fluxUnits & \\
\hline\noalign{\smallskip}
%2018-02-15 & XS1 & $92^{+30}_{-22}$ &  - &  $5.6^{+2.8}_{-2.0}$ & 62/1961\\
2018-02-28 & S1 & $76^{+6}_{-5}$ &  $<0.41$ &  $2.46^{+0.30}_{-0.29}$ & 39/50\\
2018-03-07 & S2 & $80\pm{8}$ &  $0.44^{+0.28}_{-0.35}$ & $3.36^{+0.33}_{-0.31}$ & 89/66 \\
2018-03-12 & S3 & $<88$ &  $1.88^{+0.50}_{-0.30}$ & $0.99^{+0.32}_{-0.28}$ &24/24\\
2018-03-19 & S4 & $77^{+27}_{-46}$ &  $1.89^{+0.56}_{-0.49}$ & $1.50^{+0.23}_{-0.20}$ &76/81\\
2018-03-26 & S5 & $<98$ &  $2.39^{+0.40}_{-0.66}$ & $3.20^{+0.65}_{-0.57}$ &65/65\\
2018-03-29 & S6 & $<72$ &  $2.86^{+0.51}_{-0.47}$ & $2.48^{+0.75}_{-0.61}$ &34/34\\
2018-04-07 & S7 & $<91$ &  $2.32^{+0.53}_{-0.92}$ & $3.51^{+1.01}_{-1.02}$ &37/29\\ 
2018-05-18 & S8 & $<98$ &  $2.65^{+0.41}_{-0.73}$ & $1.98^{+0.42}_{-0.39}$ &43/63\\
2018-05-23 & S9 & $<55$ &  $2.79^{+0.20}_{-0.20}$ & $2.28^{+0.28}_{-0.25}$ &117/120\\
2018-05-30 & S10 & $<59$ &  $3.22^{+0.22}_{-0.21}$ & $2.25^{+0.26}_{-0.24}$ &135/140\\
2018-06-06 & S11 & $94^{+37}_{-50}$ &  $2.54^{+0.56}_{-0.56}$ & $1.20^{+0.18}_{-0.16}$ &84/95\\
2018-06-13 & S12 & $<163$ &  $3.1^{+1.0}_{-1.7}$ & $0.49^{+0.19}_{-0.14}$ &22/24\\
2018-07-17 & X1 & $50^{+9}_{-10}$ &  $2.52^{+0.15}_{-0.15}$ & $0.27^{+0.01}_{-0.01}$ &147/145 \\
2018-09-02$^{c}$ & S13-S16 & 50$^{d}$ &  $2.1^{+1.6}_{-1.3}$ & $0.019^{+0.022}_{-0.013}$ &4.8/5\\
2019-01-10 & X2 & 50$^{d}$  &  $3.61^{+1.35}_{-0.99}$ & $0.0042^{+0.0014}_{-0.0013}$ &123/133\\
2020-01-26 & eR1 & 50$^{d}$ & $2.2^{+0.4}_{-0.5}$ & $0.49^{+0.09}_{-0.10}$ &50/55\\
2020-07-28 & eR2 & 50$^{d}$ & $3.8^{+1.2}_{-1.9}$ & $0.17^{+0.05}_{-0.06}$ &22/25\\
2022-01-26 & eR5 & 50$^{d}$ & $2.7^{+0.5}_{-0.7}$ & $0.51^{+0.10}_{-0.12}$ &41/45\\
\hline
\end{tabular}
\\
\hfill{}
\\
\tablefoot{Spectral fits to observations of \msrcnamens, modelled by {\em TBABS*COMPBB}, consisting of a single-temperature black-body Comptonised by electrons with $KT_{e}$ fixed at 5 keV, absorbed by the Galactic column of $N_{H}=6.7\times10^{20}$ cm$^{-2}$. Fits used cstat statistics, quoted errors and upper limits are 90\% confidence.\\
\tablefoottext{a}{Observed black-body temperature returned by the fit. Note that this does not represent the physical temperature of
the disc, whose emission can be better conceived as the sum of a set of black-bodies of different temperatures 
subject to multiple correction factors \citep{Mummery21}.}
\tablefoottext{b}{Optical depth of a 5 keV plasma.}
\tablefoottext{c}{Combination of observations S13--S16 (2018-08-23 -- 2018-09-11).}
\tablefoottext{d}{Black-body temperature fixed to $kT=50$ eV.}
} % end of tablefoot
} % end of small
\end{center}
\end{table*}
%\end{center}

To further probe the intrinsic disc parameters we used a physically-motivated disc model {\em TDEDISCSPEC} 
\citep{Mummery21}.
This models the disc as the sum of emission from a set of concentric regions, each with their own temperature and area, and with the intrinsic emission colour-corrected 
to model disc effects such as opacity \citep{Mummery23}. It 
has two main parameters; $T_{P}$, the intrinsic peak disc temperature before colour-correction effects are
applied, and $R_{P}$, the disc radius where the peak temperature is emitted. 
We applied this model to the first \swift-XRT
observation (S1), which is the only good-quality spectrum we have which is purely thermal.
This fits the spectrum well ($C_{r}=40/50$) giving a best-fit radius $R_{P}=6.1^{+3.3}_{-1.5}\times10^{11}$ cm and peak temperature $T_{P}=6.5^{+1.0}_{-0.6}\times10^{5}$ K ($56^{+9}_{-5}$ eV), when we fix the
index ($\gamma=0.5$) and the luminosity distance to $D_{L}=190$ Mpc.
For the second \swift-XRT
observation (S2), to model the Comptonisation we adopted the {\em SIMPL} model \citep{Steiner09}, which links an arbitrary seed spectrum to an output power-law. This has three parameters, a power-law slope
$\Gamma$, the fraction of seed photons which are scattered $f_{c}$ and a flag which we set 
so that only upscattering of photons is considered. This gave a similar quality fit to the
{\em COMPBB} model with $C_{r}=89/65$, $R_{P}=5.4^{+1.5}_{-1.2}\times10^{11}$ cm 
and peak temperature $T_{P}=7.3^{+0.6}_{-0.6}\times10^{5}$ K and $f_{c}<0.05$.
From observation S3 onwards the Comptonisation component becomes large and we find that fits with
the  {\em TBABS*SIMPL*TDEDISCSPEC} model return values of  $R_{P}$ and $T_{P}$ which are unconstrained. 
A similar effect was noted by \citet{Guolo23}, who found that such fits become unreliable when $f_{c}>0.2$.

To break the degeneracy we use the relation found in \citet{Mummery23} between
$R_{P}$ and black hole mass, 
$R_{P,12}=M_{BH,6}/4.9$, where $R_{P,12}$ is the radius of 
the peak temperature in units of $10^{12}$cm and $M_{BH,6}$ the mass of the black hole
in units of $10^{6}$\msolarns, to fix $R_{P}$. Using a variety of methods we estimate $M_{BH}$ for
\msrcname to be $4\pm{2}\times10^{6}$\msolar (see appendix \ref{Sec:mbh}).
From this we get $R_{P,12}=0.82\pm{0.41}$. Fits were conducted, using these values of $R_{P}$
and are given in Tab.~\ref{tab:discfits}. 

%\newpage
\begin{table*}[t!]
\begin{center}
\small{

\caption{Fits of Comptonised thermal disc emission to \xmm and \swiftns-XRT observations of \msrcnamens}
\label{tab:discfits}  

\begin{tabular}{llllllll}
\hline\hline                 % inserts double horizontal lines
Date & Key & $R_{P}^{a}$ & $T_{P}^{b}$ & $\Gamma^{c}$ & $f_{c}^{d}$ &  $L_{bol}^{e}$ & C/dof \\
     &     & $10^{12}$ cm & $10^{5}$ K &        &      & $10^{43}$\lumUnits & \\     
\hline\noalign{\smallskip}
2018-02-28 & S1 & 0.82 & $5.96^{+0.10}_{-0.10}$ & - & $<0.022$ &  59.8 & 42/50\\
           &    & 0.41  & $7.33^{+0.14}_{-0.10}$ & - & $<0.01$ & 34.2 & 45/50 \\
           &    & 1.23  & $5.27^{+0.11}_{-0.0}$ & - & $<0.044$ & 83.2 & 52/50 \\
2018-03-07 & S2 & 0.82 &  $6.20^{+0.12}_{-0.12}$ & $>2.67$ & $0.048^{+0.028}_{-0.044}$ & 71.9 & 95/66 \\
           &   & 0.41  & $7.79^{+0.12}_{-0.12}$ & - & $<0.0002$ &  44.2 & 89/66 \\
           &    & 1.23  & $5.35^{+0.13}_{-0.13}$ & $>4.5$ & $0.13$ &  93.6 & 107/66 \\
2018-03-12 & S3 & 0.82 &  $4.50^{+0.50}_{-0.85}$ & $>2.1$ & $0.19^{+0.81}_{-0.18}$ &  23.0 &25/24\\
            &     & 0.41  & $5.81^{+0.50}_{1.00}$ & $3.2^{+1.8}_{-1.8}$ & $0.11$ & 15.4 & 25/24 \\
            &     & 1.23  & $3.91^{+0.52}_{-0.76}$ & $3.8^{+1.2}_{-1.4}$ & $0.22$ & 29.8 & 25/24 \\
2018-03-19 & S4 & 0.82 &  $4.13^{+0.11}_{-0.20}$ & $3.8^{+0.2}_{-0.3}$ & $1.0^{+0}_{-0.61}$ & 25.6 &77/81 \\
           &     & 0.41  & $5.47^{+0.14}_{-0.20}$ & $4.1^{+0.2}_{-0.6}$ & $1.0$ &  18.8 & 76/81 \\
           &     & 1.23  & $3.53^{+0.08}_{-0.19}$ & $3.8^{+0.2}_{-0.3}$ & $1.0$ & 31.2 & 78/81 \\
2018-03-26 & S5 & 0.82 &  $4.62^{+1.12}_{-0.30}$ & $3.7^{+0.3}_{-0.6}$ &  $1.0^{+0}_{-0.77}$ &  42.4 &65/65 \\
           &     & 0.41  & $6.55^{+0.75}_{-0.30}$ & $3.9^{+0.3}_{-1.0}$ &  $0.5$ &  34.2 & 65/65 \\
           &     & 1.23  & $4.23^{+0.55}_{-0.20}$ & $3.5^{+0.4}_{-0.4}$ & $0.6$ &  55.8 & 65/65 \\
2018-03-29 & S6 & 0.82 &  $5.34^{+0.5}_{-1.02}$ & $2.5^{+0.9}_{-0.9}$ &  $0.11^{+0.89}_{-0.09}$ &  48.8 &32/34 \\
           &     & 0.41  & $6.98^{+0.52}_{-0.80}$ & $2.1^{+1.0}_{-1.0}$ & $0.07$ & 36.0 & 30/34 \\
           &     & 1.23  & $4.61^{+0.49}_{-0.80}$ & $2.7^{+0.7}_{-0.8}$ & $0.12$ & 60.3 & 32/34 \\
2018-04-07 & S7 &  0.82 &  $6.12^{+0.36}_{-0.60}$ &  $1.6^{+1.8}_{-1.6}$ & $0.01^{+0.21}_{-0.005}$ &  75.3 &33/29 \\
           &     & 0.41  & $7.72^{+0.38}_{-0.40}$ & $1.2^{+1.4}_{-1.1}$ & $0.02$ &  51.0 & 30/29 \\
           &     & 1.23  & $5.29^{+0.39}_{-1.10}$ & $2.3^{+1.8}_{-2.3}$ & $0.03$ & 92.4 & 31/29 \\
2018-05-18 & S8 & 0.82 &  $4.43^{+0.74}_{-0.30}$ &  $3.3^{+0.4}_{-0.7}$ & $1.0^{+0}_{-0.89}$ &  30.8 &43/63 \\
           &     & 0.41  & $5.43^{+0.15}_{-0.40}$ & $3.5^{+0.4}_{-1.2}$ & $1$ &  21.0 & 43/63 \\
           &     & 1.23  & $3.45^{+0.08}_{-0.30}$ & $3.4^{+0.4}_{-0.8}$ & $1$ & 32.6 & 43/63 \\
2018-05-23 & S9 & 0.82 &  $4.90^{+0.34}_{-0.60}$ & $3.0^{+0.4}_{-0.4}$ &  $0.27^{+0.73}_{-0.14}$ &  38.9 &116/120 \\
           &     & 0.41  & $6.46^{+0.32}_{-0.48}$ & $2.8^{+0.4}_{-0.4}$ & $0.22$ &  29.0 & 114/120 \\
           &     & 1.23  & $4.2^{+0.33}_{-0.48}$ & $3.1^{+0.2}_{-0.4}$ & $0.29$ &  47.3 & 116/120 \\
2018-05-30 & S10 & 0.82 &  $4.45^{+0.48}_{-0.40}$ &  $2.7^{+0.3}_{-0.3}$ & $0.4^{+0.6}_{-0.22}$ &  33.7 &136/140 \\
           &     & 0.41  & $5.16^{+0.08}_{-0.20}$ & $2.7^{+0.3}_{-0.3}$ & $1$ & 22.8 & 137/140 \\
           &     & 1.23  & $3.85^{+0.28}_{-0.5}$ & $2.8^{+0.2}_{-0.3}$ & $0.36$ & 40.2 & 135/140 \\
2018-06-06 & S11 & 0.82 &  $3.43^{+0.11}_{-0.10}$ & $2.9^{+0.3}_{-0.3}$ &  $1.0^{+0}_{-0.60}$ &  17.5 &88/95 \\
           &     & 0.41  & $4.66^{+0.14}_{-0.30}$ & $3.0^{+0.3}_{-0.3}$ & $1$ &  14.1 & 87/95 \\
           &     & 1.23  & $2.89^{+0.08}_{-0.2}$ & $3.0^{+0.3}_{-0.3}$ &  $1$ &  19.9 & 88/95 \\
2018-06-13 & S12 & 0.82 &  $2.76^{+0.1}_{-0.58}$ & $2.8^{+0.6}_{-1.0}$ &  $1.0^{+0}_{-0.97}$ &  7.85 &22/24 \\
           &     & 0.41  & $3.73^{+0.22}_{-0.60}$ & $2.8^{+0.6}_{-1.2}$ & $1$ &  6.3 & 22/24 \\
           &     & 1.23  & $2.53^{+0.10}_{-0.51}$ & $2.8^{+0.6}_{-1.0}$ & $0.56$ &  10.3 & 22/24 \\
2018-07-17 & X1 & 0.82 &  $2.91^{+0.02}_{-0.03}$ & $3.7^{+0.1}_{-0.1}$ &  $1.0^{+0}_{-0.22}$ & 6.6 & 162/145 \\
           &     & 0.41  & $3.82^{+0.02}_{-0.03}$ & $3.8^{+0.1}_{-0.1}$ &$1$ & 4.8 & 154/145 \\
           &     & 1.23  & $2.49^{+0.02}_{-0.03}$ & $3.6^{+0.1}_{-0.1}$ & $1$ &  8.2 & 166/145 \\
2018-09-02$^{f}$ & S13-S16 & 0.82 &  $1.82^{+0.33}_{-0.80}$ & $3.8^{+1.2}_{-1.6}$ & $1.0^{+0.0}_{-0.99}$ &  0.99 &5/4 \\
           &     & 0.41  & $2.28^{+2.23}_{-1.0}$ & $3.7^{+1.3}_{-1.5}$ &$1$ & 0.63 & 5/4 \\
           &     & 1.23  & $1.57^{+0.25}_{-0.79}$ & $3.8^{+1.2}_{-1.6}$ & $1$ & 1.24 & 5/4 \\
2019-01-10 & X2 & 0.82 &  $1.47^{+0.71}_{-0.40}$ & $2.9^{+0.7}_{-0.7}$ & $0.22^{+0.78}_{-0.20}$ & 0.31 &124/132 \\
           &     & 0.41  & $1.38^{+0.08}_{-0.30}$ & $2.9^{+0.7}_{-0.6}$ & $1$ & 0.12 & 124/132 \\
           &     & 1.23  & $1.56^{+0.46}_{-0.80}$ & $2.9^{+0.7}_{-0.6}$ & $1$ &  0.70 & 124/132 \\ 
\hline
\end{tabular}
\\
\hfill{}
\\
}
\end{center}
\tablefoot{Spectral fits to observations of \msrcnamens, modelled by {\em TBABS*SIMPL*TDEDISCSPEC}, consisting of 
photons from a thermally-emitting disc, Compton-upscattered into a simple
power-law, absorbed by the Galactic column of $N_{H}=6.7\times10^{20}$ cm$^{-2}$. 
The {\em TDEDISCSPEC} model is used with the index $\gamma$ fixed to 0.5 for a face-on
disc with the temperature maximum inside the disc body, luminosity distance, $D_{L}=190$Mpc and normalisation=1. The photon power-law index, $\Gamma$ 
is left free in {\em SIMPL} and the model is used in upscattering-only mode.
Fits used cstat statistics, quoted errors are 90\% confidence.\\
\tablefoottext{a}{$R_{P}$ is the radius where the disc emits the peak temperature. This is fixed at values of 0.82, 0.41 and 1.23 $\times10^{12}$cm using the relation given in equation 13
of \citet{Mummery23}, corresponding to the black hole mass range adopted in appendix \ref{Sec:mbh}, $M_{BH}=4^{+2}_{-2}\times10^{6}$\msolarns.} 
\tablefoottext{b}{Peak intrinsic temperature of the disc. }
\tablefoottext{c}{Slope of the {\em SIMPL} power-law component.} 
\tablefoottext{d}{Fraction of scattered photons.}
\tablefoottext{e}{Bolometric luminosity. Due to simplifications in the treatment of photons outside of the
X-ray band by {\em TDEDISCSPEC}, {\em XSPEC} itself can not be used to calculate the 
integrated bolometric luminosity of the model. We have used equation 13 of \citet{Mummery23}
to calculate the disc luminosity from values of $R_{P}$ and $T_{P}$ and have used {\em XSPEC}
to calculate the extra luminosity contained in the upscattered photons.}
\tablefoottext{f}{Combination of observations S13--S16 (2018-08-23 -- 2018-09-11)}
} % end footnote
 % End small
%\end{center}
\end{table*}
%\end{center}

\subsection{Bolometric Lightcurve}
\begin{figure}
       \resizebox{\hsize}{!}{\includegraphics{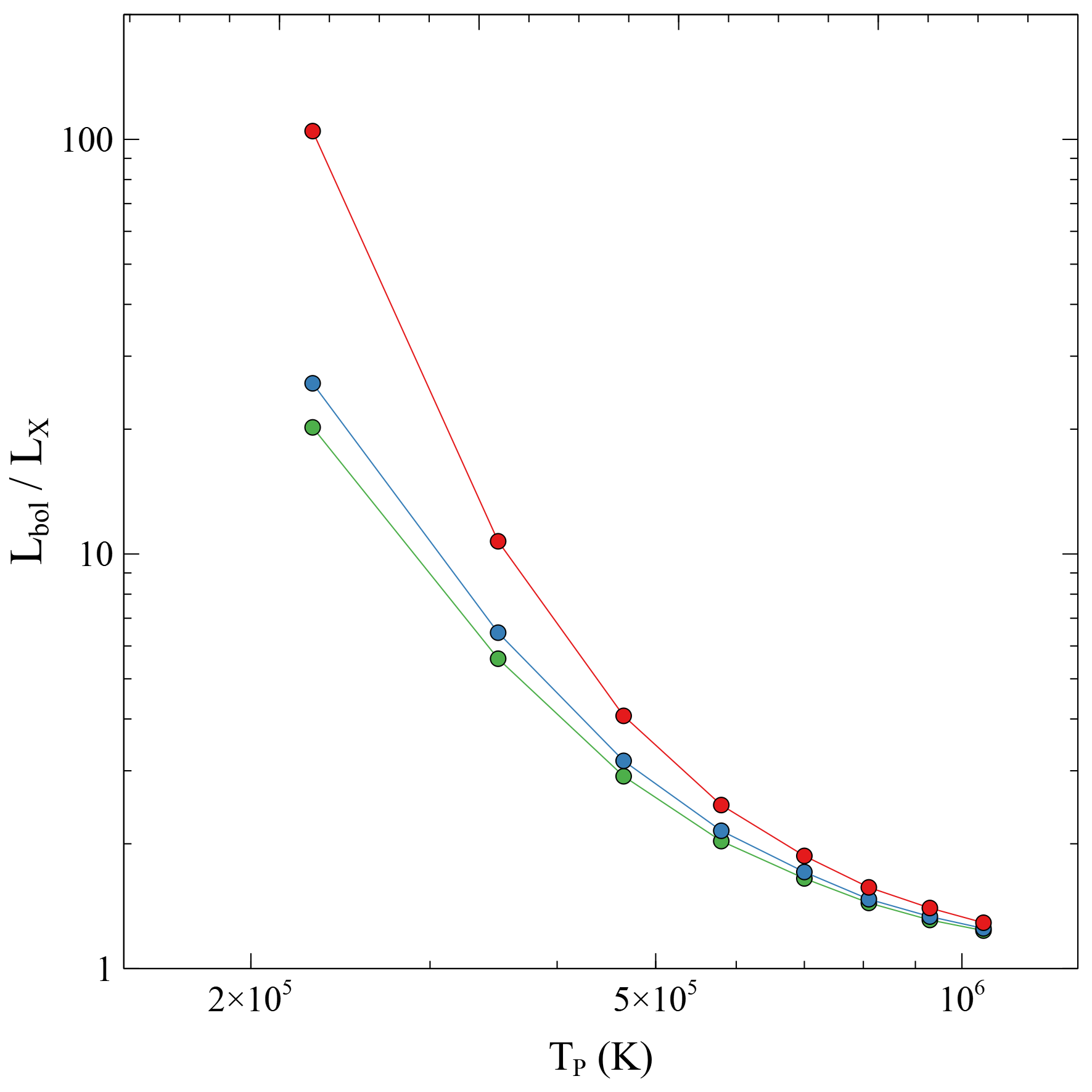}}
    \caption{Bolometric correction factor for 0.2--2 keV emission from a 
    single temperature thermal spectrum (red)
    and a Comptonised thermal spectrum with electrons of temperature 2 keV and $\tau=2$ (blue) and
    5 keV with $\tau=1$ (green), plotted against the peak temperature of the thermal component.}
    \label{fig:bol_corr}
\end{figure}

The X-ray light curve does not run parallel with the bolometric light curve and X-ray decay rates can be exaggerated by the disc temperature
declining as the luminosity falls. In fact the discrepancy becomes
exponential as the temperature drops below that of the observed X-ray energy range \citep[see figures 14 and 15 of][]{Mummery23}.
The work of \citet{Mummery23} considered the correction ($\eta_{X}$) needed to convert between thermal X-ray and bolometric luminosity. This varies slowly for $kT\geq50$ eV but rises exponentially as the temperature decreases. In section~\ref{sec:specfits} we show that the spectrum of \xsrcname becomes
modified by Compton up-scattering of disc photons. This will temper the dramatic variation of $\eta_{\rm X}$ as the luminosity and hence temperature of seed photons drops, due to lower energy photons being shifted into the X-ray band. As an example of this effect, in Fig.~\ref{fig:bol_corr} we show $\eta_{\rm X}$ v $T_{\rm X}$ for the Comptonisation model {\em COMPBB}, with electron temperature $kT_{\rm e}=2$ keV and optical depth fixed at $\tau=2$ or $kT_{\rm e}=5$ keV and $\tau=1$ (values which
are permitted by the spectral fits of section~\ref{sec:specfits}). 

%If we estimate $L_{\rm bol}$ from the X-ray spectrum using $\eta_{X}$ appropriate for a Comptonised
%black-body with $kT_{\rm e}=5$ keV and $\tau=1$ and assume a disc temperature dependency, 
%appropriate for an accretion disc, of $T_{d}\propto L_{\rm bol}^{1/4}$ for all of the
%observations then the decay index flattens to $t^{-3.6\pm{0.2}}$ (Fig.~\ref{fig:Xlc_opt}).

\begin{figure}
  %\resizebox{\hsize}{!}{\includegraphics{X_opt_rad_justflare_uvw2.png}}
  \resizebox{\hsize}{!}{\includegraphics{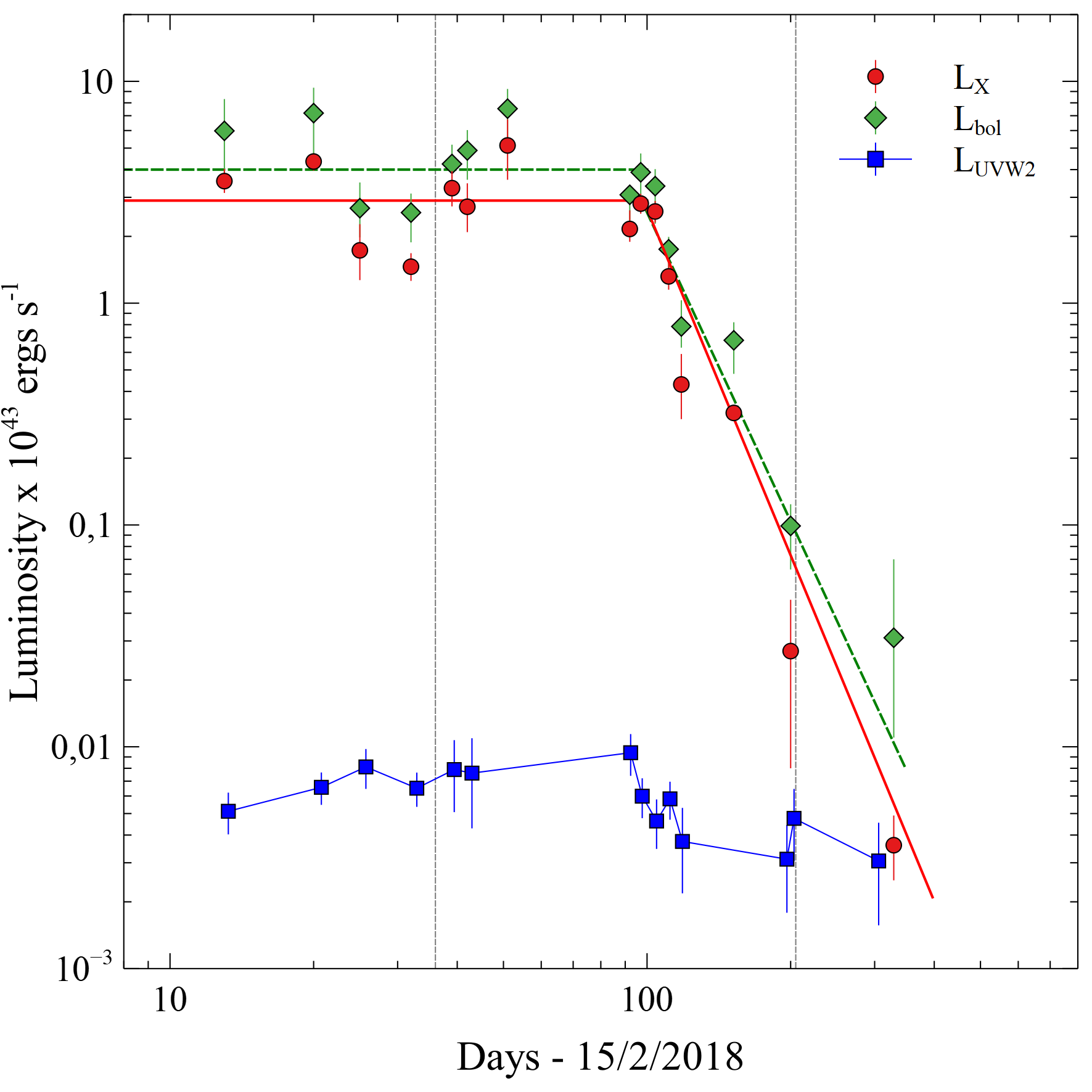}}
    \caption{X-ray (red circles), host-subtracted, absorption-corrected, 
    UVOT-UVW2 filter (blue squares) and bolometric (green diamonds) luminosity light curves. 
Bolometric luminosities have been calculated from  
%the unabsorbed X-ray flux using the correction factors of Fig.~\ref{fig:bol_corr}.
fits of a Comptonised thermal disc model to the X-ray spectra 
(see section~\ref{sec:specfits}).
Solid red line gives a fit of constant flux in the early phase and a decay of $t^{-5.2}$ from day 100
to the X-ray luminosity. Dashed green line gives the same for the bolometric luminosity with a
decay of $t^{-4.6}$. 
Grey vertical dashed lines indicate the times of the radio
non-detections.}
    \label{fig:Xlc_opt}
\end{figure}

We can estimate $L_{\rm bol}$ using fits of the Comptonised thermal disc spectrum
(Tab.~\ref{tab:discfits}) 
to the X ray spectra. In bolometric light, the steep decay of the
X-ray light curve starting from day 100, flattens to a decay index
of $t^{-4.6\pm{0.3}}$ (Fig.~\ref{fig:Xlc_opt}).

From the \xmm slew observation, the peak luminosity was $L_{\rm X}=6.5^{+3.2}_{-2.3}\times10^{43}$\lumUnitsns and 
applying the correction factor appropriate for a thermal model of kT=92 eV ($1\times10^{6}$ K), 
$L_{\rm bol}=8.3^{+4.1}_{-3.0}\times10^{43}$\lumUnitsns. The total luminosity
integrated over the bolometric light curve is $5\times10^{50}$ ergs, equivalent to an accreted mass of 
$\sim0.003$ \msolar during the 329 days of observations, for a mass to light conversion efficiency $\eta=0.1$ \citep{Gruzinov98}.
As the observations may have missed the peak of the emission this value is necessarily a lower limit.

\subsection{\nustar observation}
\begin{figure}
        \resizebox{\hsize}{!}{\includegraphics{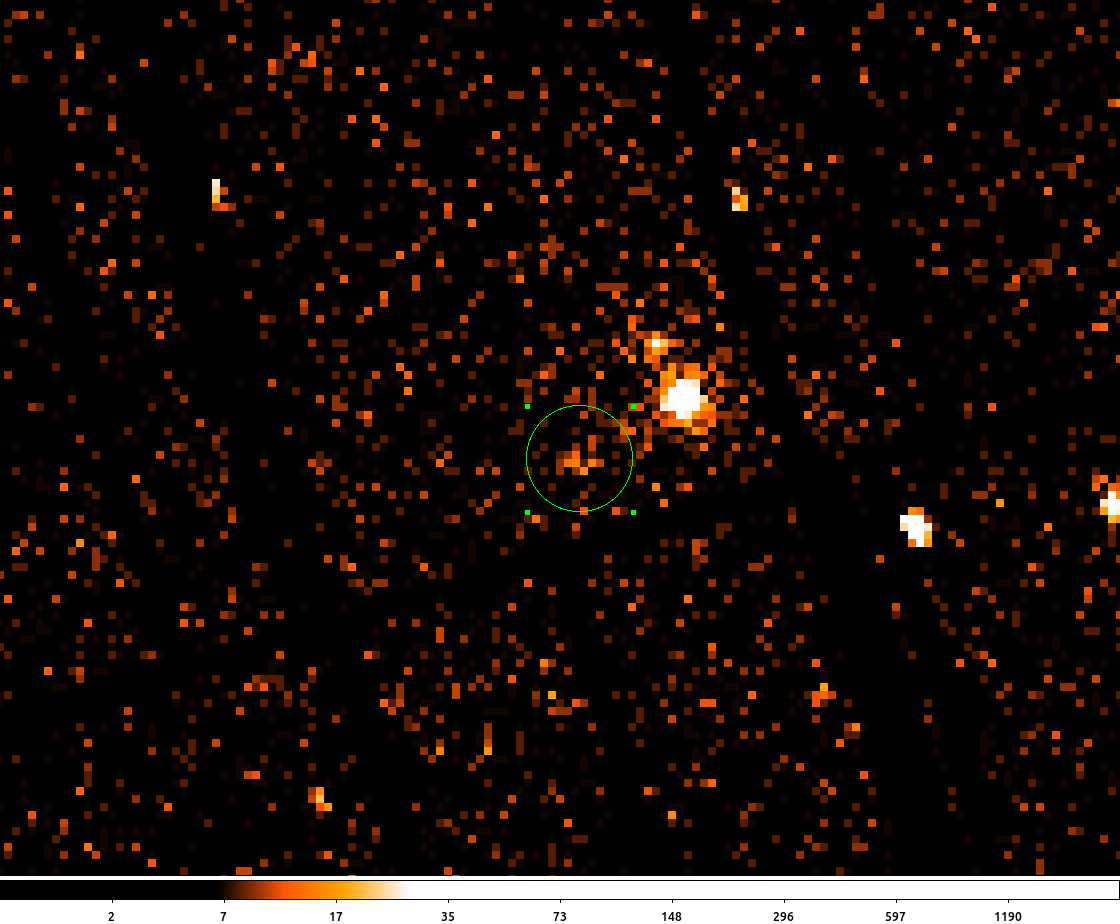}}
        \resizebox{\hsize}{!}{\includegraphics{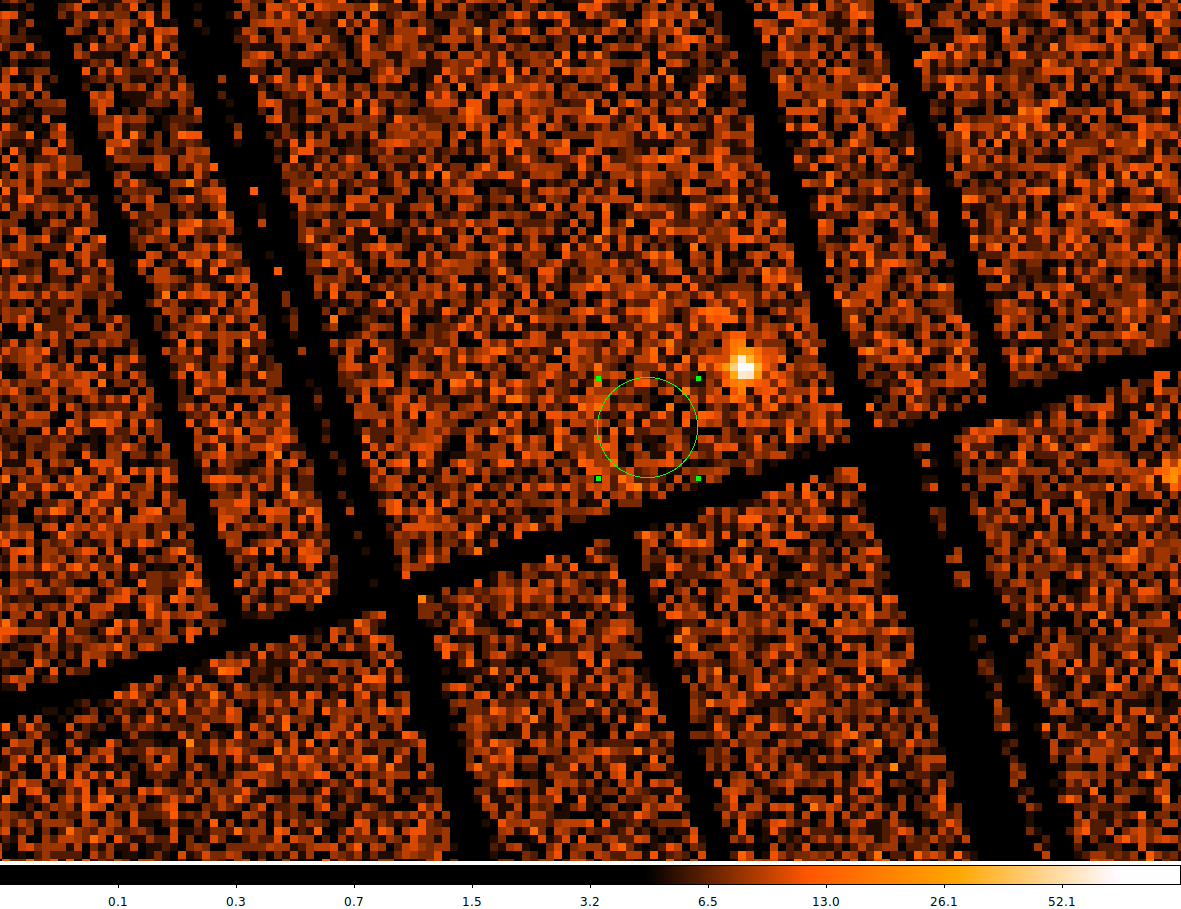}}
        \resizebox{\hsize}{!}{\includegraphics{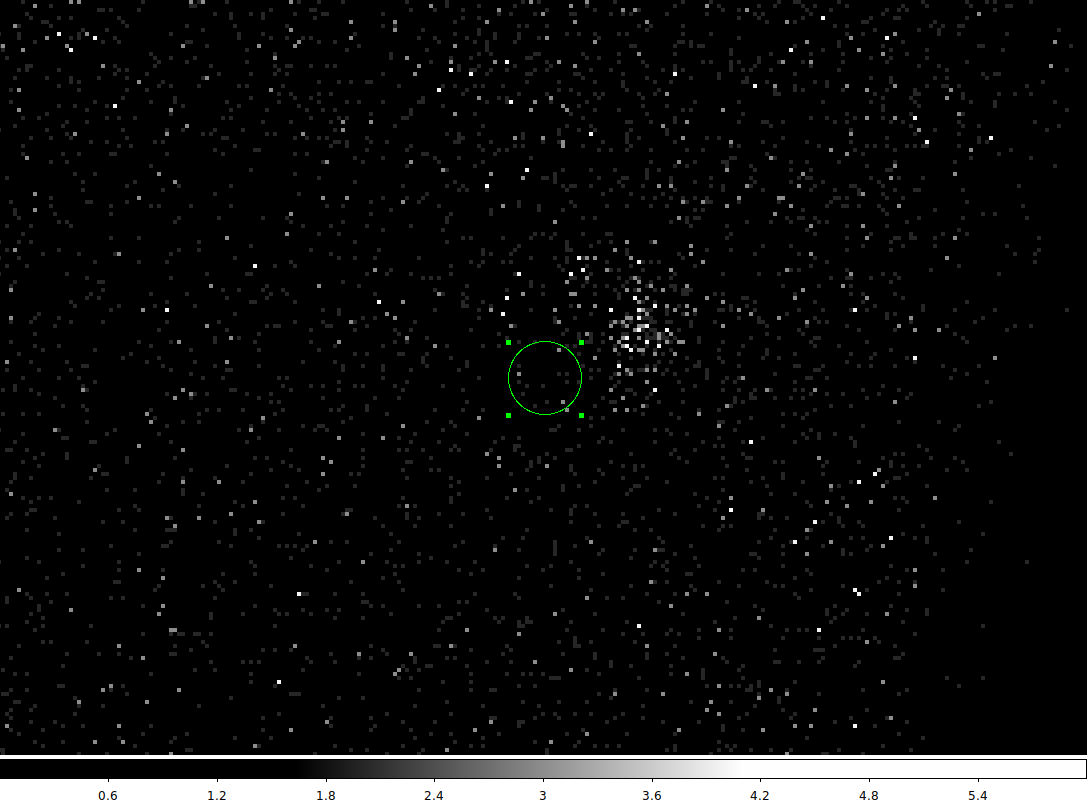}}
    \caption{Upper: EPIC-pn, 0.2-2 keV image of \xsrcname taken on 2019-01-10, Middle: Epic-pn, 2-10 keV image, Lower: NuSTAR, FPA, 3-100 keV image centred on the position of \xsrcnamens. The source, whose position is denoted by a green circle, is not detected in the latter two images.}
    \label{fig:xmmNustarObs}
\end{figure}

To help determine whether the steep flux decay starting around day 100, was intrinsic or could have been caused by local absorption, a 69\,ks observation was made with \nustar \citep{Harrison13} under ObsID  90401640002 simultaneously with the second XMM-Newton pointed observation. The \nustar data were reduced by using the standard \texttt{NuSTARDAS} software v1.8.0 and CALDB 20170727. We used  a circular extraction region with a radius of $25''$ centered on the coordinates of \xsrcname to extract any source counts. Background counts were extracted from a source free region in the same quadrant, using a circular region with a radius of $100''$. We used the same regions for both telescopes, FPMA and FPMB.
The source was barely detected in the \xmm soft (0.2--2 keV) band and was not detected at all with
\nustar with an upper limit of $2.4\times10^{-3}$c/s from the combined FPMA and FPMB data in the 3--20\,keV band, equating to a 2-$\sigma$ upper limit of $F_{X}<4.8\times10^{-14}$\fluxUnits in that band\footnote{Calculated using a source circle of radius 29" \citep[containing 50\% of the counts;][]{Harrison13},
a local background and a spectrum of a power-law of $\Gamma=3.4$ absorbed by the Galactic column.} (Fig.~\ref{fig:xmmNustarObs}). 
%This makes it unlikely that the flux reduction is due to an increase in absorption.
Using the combined \xmm and \nustar spectrum we can set constraints on the level of absorption which would be needed to explain the drop in flux purely by intrinsic absorption in the vicinity of the source. If we fit that combined spectrum with the spectral parameters from the {\it first} \xmm pointed observation then to get a good fit we would need to add a combination of a neutral absorber with $N_{\rm H}=3.5^{+2.5}_{-2.0}\times10^{21}$cm$^{-2}$ and an ionised absorber \citep[ZXIPCF; ][]{Reeves08} with
$N_{\rm H}=3.8^{+2.4}_{-1.8}\times10^{23}$cm$^{-2}$ and ionisation parameter $\log \xi=2.1^{+0.8}_{-0.2}$
both at the redshift of the source, yielding $C_{r}=122/133$. A single absorber does not provide a good fit to the data.

\subsection{eROSITA and late-time Swift observations}

The position of \xsrcname was observed during five eROSITA \citep{Predehl21} all-sky surveys (eRASS) between 2020 and 2022. The eROSITA data were calibrated and cleaned using the pipeline version 020 of the eROSITA Science Analysis Software (eSASS, \citealt{brunner_etal2022}). For each eRASS, we merged all photon events from the seven telescopes into one event list file. The X-ray spectra and light curves for each eRASS were extracted using the \textsc{srctool} in eSASS (version \texttt{20211004}). A circular region with a radius of $40''$ was chosen as the source region for all eRASSs. While a source-free annular region with an inner radius of $100''$ and an outer radius of $250''$ was selected as the background region.

\begin{figure}
         \resizebox{\hsize}{!}{\includegraphics{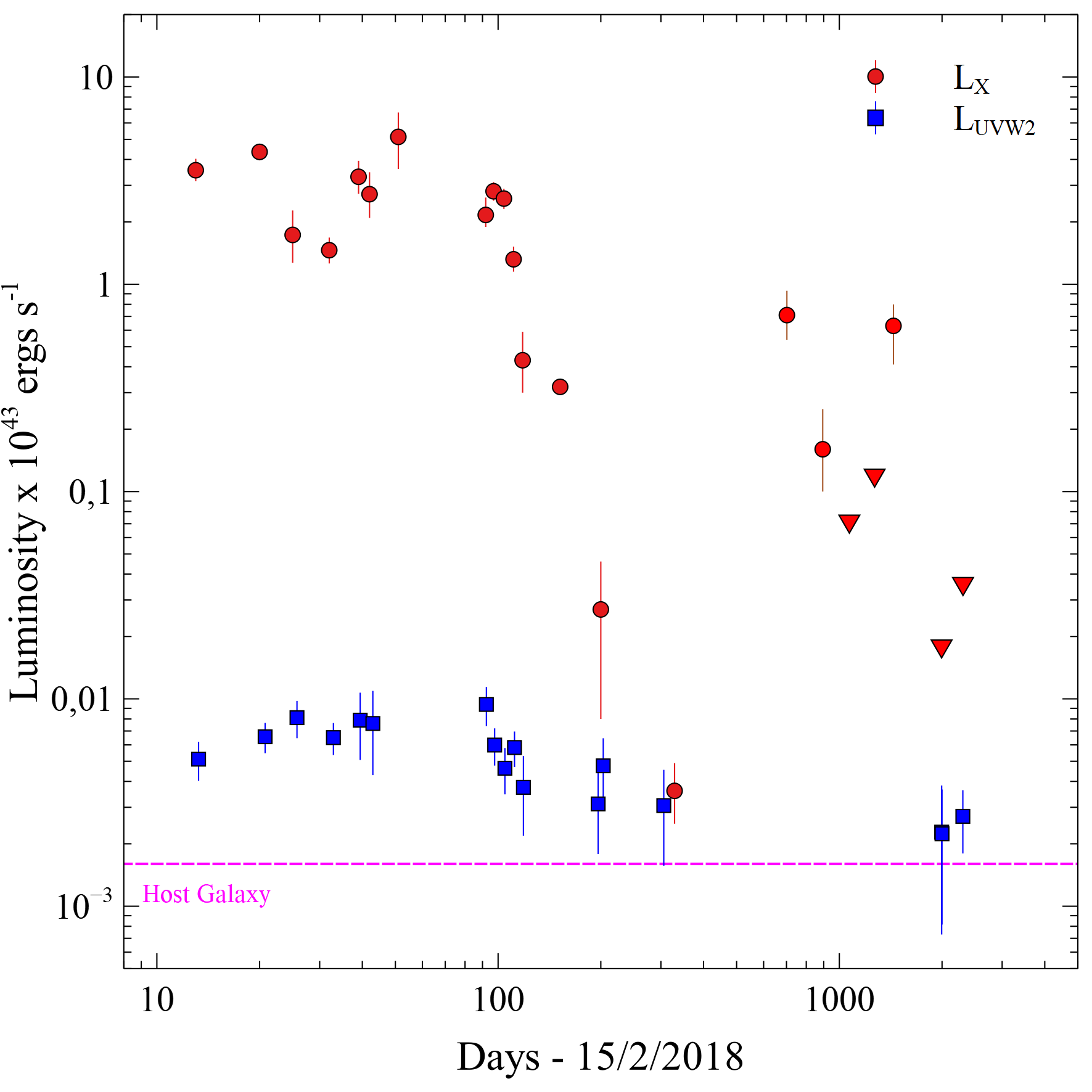}}
        \resizebox{\hsize}{!}{\includegraphics{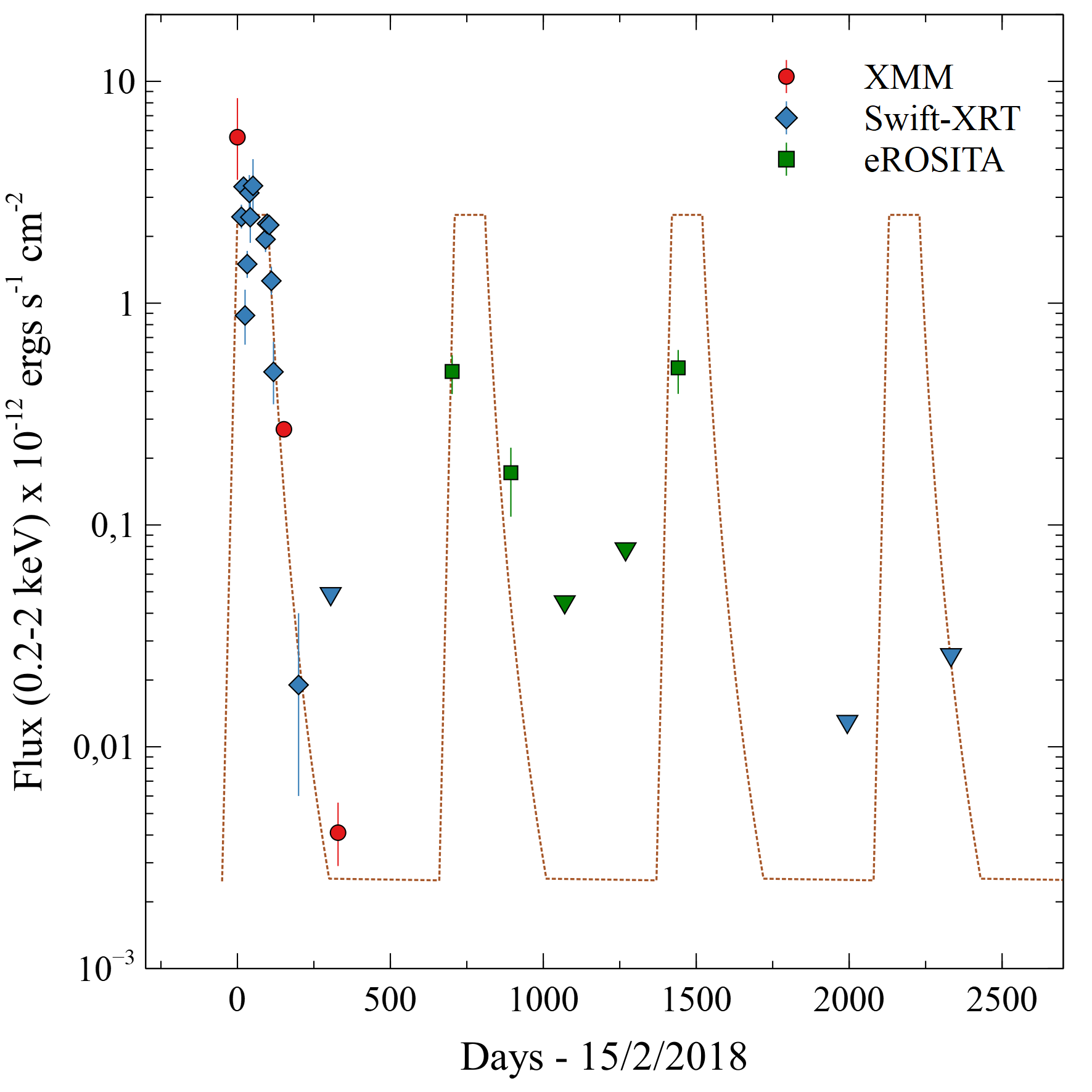}}
    \caption{Upper: X-ray (red circles) and host-subtracted, Galactic absorption-corrected, UVOT-UVW2 filter (blue squares) luminosity light curves. The modelled host galaxy UVW2 luminosity is shown for reference (pink dashed line), Lower: Full X-ray observed flux light curve calculated using the {\em COMPBB} model (see Tab.~\ref{tab:specfits}). The dotted line represents a flare consisting of  a 50 day rise to a 100 day plateau of $2.5\times10^{-12}$
    \fluxUnits and a 200 day decay with index $t^{-5}$, repeating every 710 days. Downward triangles denote 2-sigma upper limits.}
    \label{fig:fullLightCurve}
\end{figure}

\xsrcname was detected by eROSITA in eRASS1, 2 and 5 (Tab.~\ref{tab:xobs}) at fluxes above the low state of 2019-01-10. 
The full light curve is shown in Fig.~\ref{fig:fullLightCurve}, and although the monitoring is
very sparse, the data are consistent with three roughly equally-spaced flares. This is very similar to the long-term light curve of \eroszerofour, which showed multiple flares interpreted as repeated partial disruptions of a stellar object \citep{Liu23,Liu24}.
A suggested three-disruption light curve, consisting of a 50-day
rise, a 100-day plateau and a 200-day decay, repeating every 710 days, has been overlaid on the figure. In this scenario, the non-detections from 
\swift on days 1994 and 2305 were either unfortunately placed 
and just missed the fourth flare, or the star has now been
totally destroyed. Future X-ray observations around February 2026 should be able to confirm this.

\section{Discussion}
From the lack of broad and narrow emission lines in the optical spectrum, the initially purely thermal X-ray spectrum and the very low luminosity of the second \xmm pointed observation, we strongly disfavour AGN activity.
The Wise colours, W1=13.476 mag (Vega), W2=13.445 mag (Vega) also support the classification of \msrcnamelong as a non-active galaxy \citep{stern12}. Therefore, we ascribe the observations of \xsrcname to a new accretion event, i.e. a TDE.
%As the thermal emission closely follows the expected $L\propto T^{4}$ relation (Fig.~\ref{fig:LvT4}) then it also seems clear that the X-ray radiation comes from the inner part of an accretion disc, rather than from external shocks \citep[e.g.][]{Liu:2019a,Steinberg22}.
The optical/UV characteristics of \xsrcname are similar to many other TDEs, given the rather poor
monitoring data that we have, i.e. a standard temperature of a few $x10^{4}$ K and a slow decay. However, the long-term 
ASAS-SN light curve does not show an optical peak at any stage (Fig.~\ref{fig:asassn}), which is unusual for a TDE\footnote{Although an 
increasing number of X-ray discovered TDEs are being found with no simultaneous or preceding optical transient \citep{Sazonov21,Jin24}.}. Point to point variations in the V-band light curve yield an upper limit of $3\times10^{42}$\lumUnits for any possible optical peak prior to the X-ray discovery. Optical emission
from \xsrcname is most easily understood as emission from the outer regions of the accretion disc
\citep[e.g.][]{MummeryBalbus20, MummeryVV24, GuoloMumm25}.

We can then attempt to interpret the observations in terms of a TDE model capable of addressing the distinctive X-ray characteristics of \xsrcnamens:
1. A plateau of at least 100 days.
2. The development of a harder component on a timescale of $\sim10$ days.
3. A steep decay of the X-ray flare by a factor $\sim500$ over $\sim200$ days after the end of the plateau.
4. Repeated flares in later data.

\subsection{X-ray light-curve plateau} 

The first TDEs discovered by ROSAT showed sharp decays in their light curves interpreted as  
a response to a reduction in the rate of returning debris 
\citep[][ and references therein]{Komossa:2004b}. 
As can be seen from Fig.~\ref{fig:Xlc_opt}, the X-ray flux, and the inferred bolometric flux in 
\xsrcname vary
little over the first 100 days of observations. We interpret this as a plateau similar to that seen
in several sources such as XMMSL2~J144605.0+685735 \citep{Saxton:2019a} and ASSASN-15oi \citep{Gezari17}. 
The lack of previous X-ray observations prevent us from knowing the full duration of the plateau \footnote{We note that had the Einstein Probe \citep{Yuan:2016a}
been in nominal operations prior to 2018, the question of previous behaviour would have been resolved, illustrating what a great asset that mission will be to TDE astronomy.}.
%It may be a coincidence that the source spectrum changed soon after being discovered and the source may have been in the thermal state for a long time
There are two main reasons why the light curve might decay more slowly than the 
rate of incoming material. When the fall back rate is super-Eddington, a combination of energy advected into the black hole (photon trapping) \citep{Abramowicz88,Ohsuga02}, 
increasing colour-correction factor \citep{Wen20} and 
loss of material in high-speed winds \citep[e.g.][]{Dai18} leads to a self-limiting emitted
X-ray luminosity, which has been used to explain the flat light curve of 3XMM~J150052.0+015452 \citep{Lin_1500,Cao23}.
The fact that an accretion disc state change (i.e. development of a hard component), which 
is expected to happen at sub-Eddington accretion rates \citep[][and see 
Section~\ref{Sec:coronacreation}]{Esin97,Maccarone03}, occurred in \xsrcname 
during the plateau phase argues against a sustained period of super-Eddington accretion during the plateau in this case.

A more likely explanation is that debris spreads out into an accretion disc which accretes viscously \citep{Cannizzo90,MummeryBalbus20,Jonker20}.
If this happens then the disc can act as a reservoir of accretable material until the rate of incoming debris
falls below the rate of material being consumed centrally, in which case the luminosity will drop again. 

The UV/optical luminosity during the plateau is lower than seen in many 
TDEs, with a mean plateau luminosity in the UVOT-U filter of $L_{U}\sim5\times10^{40}$ \lumUnitsns. From the relationship between  optical plateau luminosity and black hole
mass derived in \citet{Mummery24} we obtain a mass of $3.5^{+3.5}_{-1.7}\times10^{5}$ \msolarns.
This is lower than our other estimates (Appendix~\ref{Sec:mbh}),
but we note that the scatter of the \citet{Mummery24} relationship is particularly large at the low mass end, 
while being constrained at the other end by the Hills mass.
% Could this be due to a smaller disc in turn due to the smaller disc mass from the partial TDE?

\subsection{What triggers the development of the corona and when does it happen?}

At the time of the development of the Comptonisation zone the observed temperature of the disc emission was $kT\sim80$ eV. If we take 2018-03-07 as the date when the
zone formed then, from Tab.~\ref{tab:discfits}, the bolometric luminosity was $L_{\rm bol}=7\pm{3}\times10^{43}$ \lumUnits
%using the correction factors from Fig.~\ref{fig:bol_corr},
%the bolometric luminosity is $L_{\rm bol}=4.3\times10^{43}$ \lumUnits 
and the
Eddington ratio $0.13^{+0.14}_{-0.04}$ for the adopted mass of $M_{BH}=4\pm{2}\times10^{6}$\msolar. From \citet{Mummery23} we see that disc orientation 
can increase the bolometric correction, and hence implied accretion rate, at this temperature by up to a factor 4 due to the X-ray emitting region being obscured
by the outer disc. The absence of intrinsic absorption in the X-ray spectrum makes it unlikely that we are seeing the emission at very high inclination angles so this effect is likely to be moderate;
this argument could be made yet more strongly for AT2021ehb by \citet{Yao22} where the disc was thicker.

State transitions are commonly seen in accreting Galactic black hole binaries \citep{Belloni97a,RemMc06}. Changes on timescales of days or longer in these systems have been attributed to disc instabilities \citep[e.g.][]{Dubus01, Kalemci2013}, which when scaled up to the size of SMBH are far longer than those seen here. However, in specific binaries that show signs of obscuration and are accreting near the Eddington limit (such as GRS 1915+105, V4641 Sgr, V404 Cyg), analogous spectral transitions can occur within about 10 seconds \citep{Rao00, Maitra2006, KSA2020}. 
If scaled linearly from a 10\msolar to a $10^{6}$\msolar SMBH, these rapid viscous time-scale spectral transitions would imply a state change of $10^{6}$ seconds ($\sim 10$ days), totally consistent with that seen in \xsrcnamens.
Destruction and construction of an X-ray corona, possibly associated with a tidal disruption event, has also been seen to occur on a timescale of 10s to 100 days in a changing-look AGN \citep{Ricci20,Li24_1ES}.
Finally it is interesting to compare this event with the still faster quasi-periodic eruptions 
\citep[QPEs][]{Miniutti19,Giustini20,Arcodia21} seen in other TDE systems.
The spectral change in \xsrcname is superficially similar to that seen in e.g. GSN069, where the 
effective temperature of the emission increased from 50 eV to 120 eV. 
In GSN069 this change occurs in 2000-3000 seconds \citep{Miniutti19}, several hundred times
faster than in \xsrcname. From Fig.~\ref{fig:SimpleSpecFits} it takes 12 days to increase
the effective temperature to 120 eV. Even considering the lightest possible black hole in GSN~069 
allowed by the M-sigma relation of $3.2\times10^{5}$ \msolar \citep{Wevers22}, scaling down linearly 
would still need $\sim1$ day for the change, a factor 50 slower than the QPEs. Hence it seems 
unlikely that the same mechanism is responsible for the spectral changes seen in both \xsrcname
and QPEs.

\begin{figure}
        \resizebox{\hsize}{!}{\includegraphics{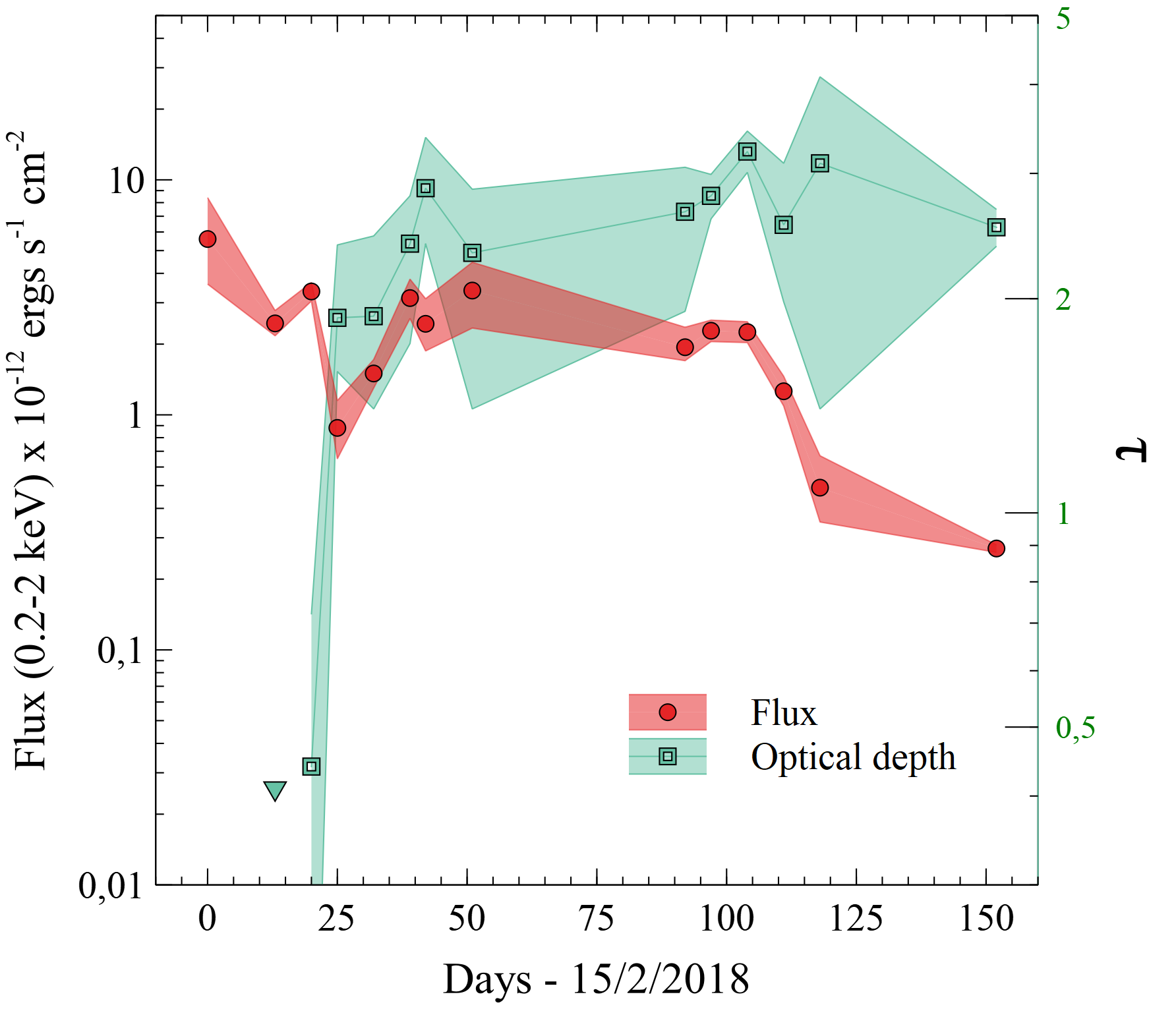}}
    \caption{Optical depth and observed 0.2--2 keV flux for the 
    first 152 days of observations of \xsrcname using a spectral model
    of {\em COMPBB} with kT$_{e}=5$keV. The shaded areas represent 90\% confidence
    intervals on the measured parameters.}
    \label{fig:fluxandtau}
\end{figure}

When the corona in \xsrcname begins to form there is a coincident drop and recovery in flux (Fig.~\ref{fig:fluxandtau})
equivalent to a loss in total bolometric luminosity of $\sim10^{49}$ ergs or accretion 
energy of $\sim10^{50}$ ergs, assuming a mass to 
light conversion factor of 0.1. If the two are related then this may suggest that 
accretion energy is temporarily redirected into the creation of the warm corona, e.g.
by magnetically heating the outer skin of the
disc \citep{Gronk20}, or into increasing its optical depth. 
Our observations show a timescale for the creation of the warm 
corona around a $M_{BH}\sim10^{6}$\msolar black hole of $\sim 10$ days which models will 
need to comply with. Similar model constraints for a transition in the opposite direction
are provided by the
destruction, or at least significant reduction, of the hot corona in $\sim 3$ days
in AT2021ehb \citep{Yao22}. 

\subsection{Why is corona creation in TDEs not ubiquitous?}
\label{Sec:coronacreation}

\citet{Guolo23} looked at X-ray data from optically-discovered TDEs, finding that at least 40\% had
X-ray emission and showing that the perceived preponderance of optical TDEs 
\citep{Gezari21,Hammerstein23} is 
strongly biased by the current superiority of optical surveys in identifying TDEs.
Only 3 of the 17 optically-selected, X-ray detected TDEs were seen to develop a hard X-ray corona.
Conversely, from a literature search we find that 10 of the 12 TDEs and candidate TDEs discovered in the \xmm slew survey had or subsequently developed a significant hard component on top 
of the thermal disc emission in at least one observation \footnote{Curiously the two \xmm slew TDE which are purely or mainly thermal are the only ones which exhibit QPEs \citep{Miniutti19,Chak21}. This may be related to the fact that QPEs, by their nature, are more easily detected as enhancements on top of very soft X-ray spectra but it does open the possibility that the mechanism responsible for QPEs is present in many TDEs.}
(Tab.~\ref{tab:slewtdes}).  
A Monte Carlo simulation ran 100,000 times gives a probability of 0.04\% that the XMM slew and optically-selected sets are drawn from the same population. If we remove from the slew sample the
TDEs which showed evidence of a possible pre-existing AGN then the ratio becomes 6 out of 7 and the probability 0.14\%.
On the face of it this suggests that there is an intrinsic difference
between optically and X-ray identified TDEs and that observed differences are not purely due to viewing angle
\citep{Dai18,Gezari21}.
However, as pointed out in \citet{Guolo23} and evident from 
Fig.~\ref{fig:bol_corr}, TDEs which develop a hard component become more luminous in X-rays, making them more likely to be detected in time-limited X-ray surveys 
such as the \xmm slew survey \footnote{Note that this will also be true for the eROSITA survey \citep{Sazonov21} and to some extent, flux-limited surveys like the Einstein Probe \citep{Yuan:2016a}.}. More importantly,
monitoring campaigns of events need to extend into epochs when the
hard component becomes detectable.
We note that two of the thermal TDEs from the \citet{Guolo23} sample,  
AT2019teq \citep{ATel15657, ATel15724} and ASASSN-15oi \citep{Hajela25} have recently developed a hard X-ray corona. Further monitoring of the optical sample is therefore needed to confirm whether the apparent discrepancy in coronal properties is real.

\citet{wevers20} proposed that the transition to a harder state occurs in TDEs when the accretion rate drops below a value of  
$f_{edd}\sim10^{-2}-10^{-1} L_{edd}$, finding that TDEs in the X-ray thermal state were more luminous than 
those displaying coronal emission ("Fainter harder brighter softer"). AT2021ehb and AT2020ocn, however, failed to show a 
simple connection between the accretion rate and X-ray hardness ratio \citep{Guolo23}.
In \xsrcname we find that the state change appears at 
$f_{edd}\sim0.1 L_{edd}$.
%and occurs during an apparent plateau in the luminosity.

We will consider three separate components while discussing the observed spectra: a thermal component from the disc,  a hot, optically-thin electron corona indicated by a power-law component with a slope less than 2.5 and a warm, optically-thick, electron corona, indicated by a power-law with a slope of 2.5 or more. 
Compton upscattering of photons from a warm corona, consisting of an electron population 
with $kT_{e}\sim0.1 - 1$ 
keV and optical depth$>1$ has been argued to be responsible 
for the characteristic shape of the X-ray soft-excess in many AGN \citep{Magdziarz98, Done:2012a, Petrucci18,Ballantyne24,Palit24}. It is envisaged as either an extra optically-
thick layer on top of the standard accretion disc \citep{Janiuk01, Gronkiewicz23} 
or a radial structure separated from the disc \citep{Done:2012a, Kubota18}.
In AGN the slope of the hot coronal power law is almost always below 2.5 \citep{Ballantyne24} except in NLS1 when they are accreting at very high rates \citep{Ojha20,Grunwald23}. We take the hot and warm 
coronae to be separate physical components as justified for example by time delays seen between energy bands \citep[e.g.][]{Gallo04}. Indeed, \citet{Liu23} recover both a hard power-law and a soft-excess component in the repeating partial TDE, \eroszerofourns.

%\smallskip

Warm coronae in AGN are found with electron temperatures between 0.1 and 1 keV \citep{Petrucci18} and optical depths of 10--20 \citep{Petrucci18, Ballantyne24}. In \xsrcname we find $kT_{e}=2.7^{+6.2}_{-2.4}$ keV and $\tau\sim3$. Similar values were found in observations of the TDEs AT2020ocn \citep[$kT_{e}=2.4^{+1.4}_{-0.7}$;][]{Cao24} and 3XMM~J150052.0+015452 \citep[$2.3^{+2.7}_{-0.8}$;][]{Cao23}, although in other observations of AT2020ocn \citep{Cao24} and in \eroszerofour \citep{Liu23} electron temperatures <1 keV, more typical of AGN, were seen.

Some events show both thermal X-ray and warm corona emission, e.g. \eroszerofour \citep{Liu23} and AT2020ocn \citep{Cao24}, who find that some spectra can be best fit with a patchy (i.e. covering factor<1 corona). The lack of visible thermal emission in persistent AGN suggests that the covering factor is always 1 there, although 
we note that this can not be proven for AGN with higher mass black holes and hence cooler discs whose thermal
emission does not reach the X-ray band. The only AGN
which do show thermal X-ray emission are those which are believed to have experienced a recent TDE, e.g. 1ES~1927+654 \citep{Ricci20}.

In summary, in some observations of TDEs the warm corona appears to be hotter, less optically thick and with a lower covering factor than the standard AGN soft excess.

The observed softening at late time of AT2021ehb and AT2018fyk
\citep{Yao22,Wevers23fyk}
can be explained by the reduction/disappearance of the hard power-law
leaving the steeper-sloped soft excess dominant but not that the source has returned to the thermal state.
\citet{Wevers21} attribute the late-time softening of AT2018fyk to a transition into the quiescent state, which
has parallels in X-ray binaries \citep{Plotkin13} and CLAGN \citep{Ruan19}.

\begin{table}
{\small
\caption{TDEs discovered in the \xmm slew survey}
\label{tab:slewtdes}      % is used to refer this table in the text
\begin{center}
\begin{tabular}{lccl}
\hline\hline 
Name & Hard   & Poss. & References \\
     & component?  &  AGN?$^{a}$ & \\
\hline 
NGC3599 & yes       &   yes  &  (1)\\
2MASX 0740  & yes  &   no & (2) \\
XMMSL1 1446+68  & yes & no & (3) \\
XMMSL2 1404-25  & yes & no & (4) \\
XMMSL1 1201+30   & yes & no & (5)\\
XMMSL1 0619-65   & yes & yes & (6)\\
SDSS J1323+4827  & yes & no & (1) \\
NGC 5092       & yes & yes  &  (7) \\
GSN 069$^{b}$       & no  &  yes &  (8)\\
XMMSL2~2030+04  & yes  &  yes  & (9) \\
MCG+07 & yes & no  &  (9) \\
2MASX 0249$^{b}$  & no  & no  & (1)\\
\hline
\end{tabular}
%\hfill{}
\end{center}
\tablebib{
(1) \citet{Esquej:2008a}; (2) \citet{Saxton:2017a}; (3) \cite{Saxton:2019a}; 
(4) This paper; (5) \cite{Saxton:2012a}; (6) \cite{Saxton:2014a}; 
(7) \cite{Li2020}; (8) \cite{Miniutti:2013a}; (9) Li et al. {\em in prep}.}\\
$^{a}$ Did the nucleus show optical emission lines indicative of a possible pre-existing AGN? \\
$^{b}$ A power law component is 
ubiquitous in GSN~069 but has very low luminosity so that the corona, if present, is extremely weak.
\cite{Chak21} did find a low-luminosity power-law component in 2MASX~J0249 in an
observation taken 16 years after discovery. However, the thermal component was still strongly
dominant and so we consider this source to have not created a prominent corona to date.
}
\end{table}

\subsection{Decay rate of the luminosity}

At the end of the plateau phase, \xsrcname dropped in X-ray (bolometric) luminosity by a factor 500 (100) in 230 days, equivalent to a decay index of 
$t^{-5.2 (-4.6)}$. This is much steeper than 
the canonical return of tidal debris ($t^{-5/3}$) \citep{Rees88}. It is also well in excess of the
steeper decline of $t^{-9/4}$ predicted from partial disruptions (pTDE) where the stellar core survives \citep{CoughlinNix19,Nixon21}, although 
simulations show that a pTDE from a {\it weak} encounter can produce a decline in returning material with index up to 5 \citep{Ryu20}.

Similar steep decays were seen in AT2018fyk \citep{Wevers21} whose X-ray luminosity
fell by a factor 5000 in 170 days, \eroszerofour with a factor 100 decay in a week \citep{Liu23} 
and RX~J133157.6-324319.7 which exhibited a $>40$-fold decrease in 17 days \citep{Malyali23}; all of these are repeating pTDEs \citep{Wevers23fyk,Liu23,Malyali23}. 
The implication is that the rapid drop to the quiescent state may have been caused by a deficit of available material in a small disc which was quickly consumed.
%If this is also the case in \xsrcname then we might expect a second flare to occur in the X-ray or optical/UV bands. Although, the ASAS-SN g band curve (fig n) does not show any activity between 2014 and 2023 we may have missed a
%flare between 2019 and 2021 due to discontinuous coverage in the higher energy  bands.

An even more abrupt decay in \swtd  has been attributed to the turn-off of
a relativistic jet due to the accretion rate falling below Eddington \citep{Zauderer:2013a}.
The deep limits we present here on radio emission from \xsrcname (section~\ref{sec:radio}) and the fact that the transition happened at an accretion rate well below Eddington make it unlikely that this is the responsible mechanism here.
%An intermediate case is IGR~J12580+0134 which had a hard X-ray spectrum
%and decayed by a factor 1000 in 500 days \citep{nikwalt12}.
%This TDE did have strong radio emission \citep{Irwin:2015a} and likely had a
%relativistic off-axis jet \citep{Lei16}. Nevertheless, \citet{Lei16} found
%that the X-ray emission was from the corona rather than the jet making it likely that this TDE shared the same decay mechanism as \xsrcname and AT2018fyk. Which begs the question of whether IGR1258 had any further flares???

The ultra-fast decline in the luminosity cycles of 
\eroszerofour may have been augmented by a change of accretion disc state \citep{Liu23}. 
Other sharp declines in AT~2019avd \citep{Wang24} which fell
by a factor $\sim100$ in $\sim50$ days while the X-ray spectrum hardened and
AT2021ehb, which dropped by a factor 10 in flux over 3 days, accompanied by a softening of the X-ray spectrum, 
reinforce the idea of a transition to a less radiatively-efficient disc state accelerating the decline.

The discovery of more flaring activity by eROSITA (Fig.~\ref{fig:fullLightCurve}) leads to a viable interpretation for the rapid decay in the light curve of \xsrcnamens,
during 2018 and 2019, as being due to a partial disruption transferring a small
quantity of material into an accretion disc.
 The fall in luminosity perhaps being accelerated by the disc transiting into a less radiatively-efficient state 
\citep{shenMatzner14}, when the accretion rate fell below a few \% of Eddington.

We note that an alternative explanation for the repeated flares could be given by thermal instabilities in a disc fed by a single disruption event \citep{shenMatzner14,Saxton:2015a,Piro25}. Models suggest that these can occur on timescales of months to a few years
\citep{shenMatzner14,Piro25,LinMetz24}, compatible with the light curves seen in pTDEs to date.

In summary the \xsrcname data are consistent with a TDE discovered
in the plateau phase, with X-ray and UV emission dominated by thermal
emission from the disc. The conditions in the accretion structure were
such that a corona quickly formed but faded away between days 100 and 300.
%leaving a low-luminosity UV-dominated TDE which lasted for at least another 1600 days. 
The rapid decay of the X-ray flux in this event indicates 
a relatively small total amount of accreted mass and the repeat flares seen in the eROSITA data make it probable that the event consists of a series of repeated partial  
tidal disruptions. We note that we were fortunate to have the eROSITA survey to confirm the secondary flares which would otherwise have gone unnoticed. 
To avoid this in future, any TDE which shows an unusually rapid decay should be followed sporadically for several years after the event to check for repeat flares.

\section{Conclusions}

We identify the \xmm slew transient, \xsrcnamens, as a TDE which showed an unusually rapid flux and spectral evolution.
The X-ray spectrum evolved from a low-temperature ($kT\sim80$ eV) purely thermal spectrum to having a significant hard component within $<5-12$ days, putting the strongest constraint to date on this disc state transition. After a plateau period of $\sim 100$ days, the X-ray luminosity decayed by a factor 500 over the following 230 days. We note that other sources with a similarly rapid decay also displayed
evidence of a powerful corona and suggest that the accelerated evolution of
these events is due to a partial tidal disruption, leading to a rapid decrease in available fuel and passage through different accretion disc states. 
Observations during the eROSITA survey detected a resurgence of activity in \xsrcname commensurate
with flares repeating with a period of $\sim710$ days and lending support to the idea that
\xsrcname has experienced multiple partial tidal disruptions.

We find that a significantly higher fraction 
of TDEs with a hard X-ray component are found in an X-ray discovered sample than in an optically-selected sample, pointing towards a possible intrinsic difference in their properties. To exclude selection bias further monitoring of the optical sample is needed.

We recommend that any future TDE which shows an unusually rapid decay should continue to be monitored to check for repeat flares.

\section*{Acknowledgements}

We thank the anonymous referee for constructive comments which improved the manuscript.
We would like to thank the XMM OTAC for approving this program. 
The XMM-Newton project is an ESA science mission with instruments and contributions directly funded by ESA member states and the USA (NASA).
The \xmm project is supported by the Bundesministerium f\"{u}r Wirtschaft 
und Technologie/Deutches Zentrum f\"{u}r Luft- und Raumfahrt i
(BMWI/DLR, FKZ 50 OX 0001), the Max-Planck Society and the Heidenhain-Stiftung.
We thank the \swift team for approving and performing the monitoring 
observations. This work made use of data supplied by the UK \swift Science Data Centre at 
the University of Leicester. 
We thank the \nustar project scientist for allocating DDT time to this
project. This research has made use of the NuSTAR Data Analysis Software (NuSTARDAS) jointly developed by the ASI Science Data Center (ASDC, Italy) and the California Institute of Technology (USA).
This publication makes use of VOSA, developed under the Spanish Virtual Observatory (https://svo.cab.inta-csic.es) project funded by MCIN/AEI/10.13039/501100011033/ through grant PID2020-112949GB-I00.
VOSA has been partially updated by using funding from the European Union's Horizon 2020 Research and Innovation Programme, under Grant Agreement nº 776403 (EXOPLANETS-A).
Some of the data presented in this paper were obtained from the Mikulski Archive for Space Telescopes (MAST). STScI is operated by the Association of Universities for Research in Astronomy, Inc., under NASA contract NAS5-26555. 
RS would like to thank the late Tomaso Belloni for conversations along the years which helped to form some of the ideas in this paper. 
This work is based on data from eROSITA, the soft X-ray instrument aboard SRG, a joint Russian-German science mission supported by the Russian Space Agency (Roskosmos), in the interests of the Russian Academy of Sciences represented by its Space Research Institute (IKI), and the Deutsches Zentrum für Luft- und Raumfahrt (DLR). The SRG spacecraft was built by Lavochkin Association (NPOL) and its subcontractors, and is operated by NPOL with support from the Max Planck Institute for Extraterrestrial Physics (MPE).
The development and construction of the eROSITA X-ray instrument was led by MPE, with contributions from the Dr. Karl Remeis Observatory Bamberg \& ECAP (FAU Erlangen-Nuernberg), the University of Hamburg Observatory, the Leibniz Institute for Astrophysics Potsdam (AIP), and the Institute for Astronomy and Astrophysics of the University of Tübingen, with the support of DLR and the Max Planck Society. The Argelander Institute for Astronomy of the University of Bonn and the Ludwig Maximilians Universität Munich also participated in the science preparation for eROSITA.
Some plots in this paper were produced using the 
Veusz software package (http://veusz.github.io/).
GM acknowledges support from grants n. PID2020-115325GB-C31 and n. PID2023-147338NB-C21 funded by MICIU/AEI/10.13039/50110001103. MG is funded by Spanish MICIU/AEI/10.13039/501100011033 and ERDF/EU 
grant PID2023-147338NB-C21. KA acknowledges support provided by the National Science Foundation through award AST-2307668 and from the Alfred P. Sloan Foundation.
%%%%%%%%%%%%%%%%%%%%%%%%%%%%%%%%%%%%%%%%%%%%%%%%%%
\section*{Data Availability}

The data and scripts used in this paper will be made available on Zenodo post-publication. The Zenodo doi is 10.5281/zenodo.14601857.



\bibliographystyle{aa}
\bibliography{rds_references.bib} % if your bibtex file is called example.bib

%%%%%%%%%%%%%%%%%%%%%%%%%%%%%%%%%%%%%%%%%%%%%%%%%%

%%%%%%%%%%%%%%%%% APPENDICES %%%%%%%%%%%%%%%%%%%%%

\appendix

\section{Spectral analysis of the adjacent source 2SXPS~J140442.5-251104}

Spectral products were extracted from \swift-XRT observations of the source, 2SXPS~J140442.5-251104, which lies at a distance

of 67 arcseconds from \xsrcnamens, using the 
on-line XRT data products tool. A fit was performed yielding a good fit ($\chired=54/56$) for 
a power-law of slope $\Gamma=1.66^{+0.13}_{-0.12}$, absorbed by the Galactic column
of \gnhs and a further intrinsic absorber of $N_{H,i}=2.3\pm{0.5}\times10^{21}$ cm$^{-2}$ both
modelled with the {\em TBABS} model. No appreciable variability was found in the light curve of 
the source with in-band fluxes of $F_{\rm 0.2-2 \: keV}=5.5\times10^{-14}$ \fluxUnits and $F_{\rm 2-10 \: keV}=1.9\times10^{-13}$ \fluxUnitsns.

\section{X-ray short-term light curve and black hole mass}
\label{Sec:mbh}

\begin{figure}
        \resizebox{\hsize}{!}{\includegraphics{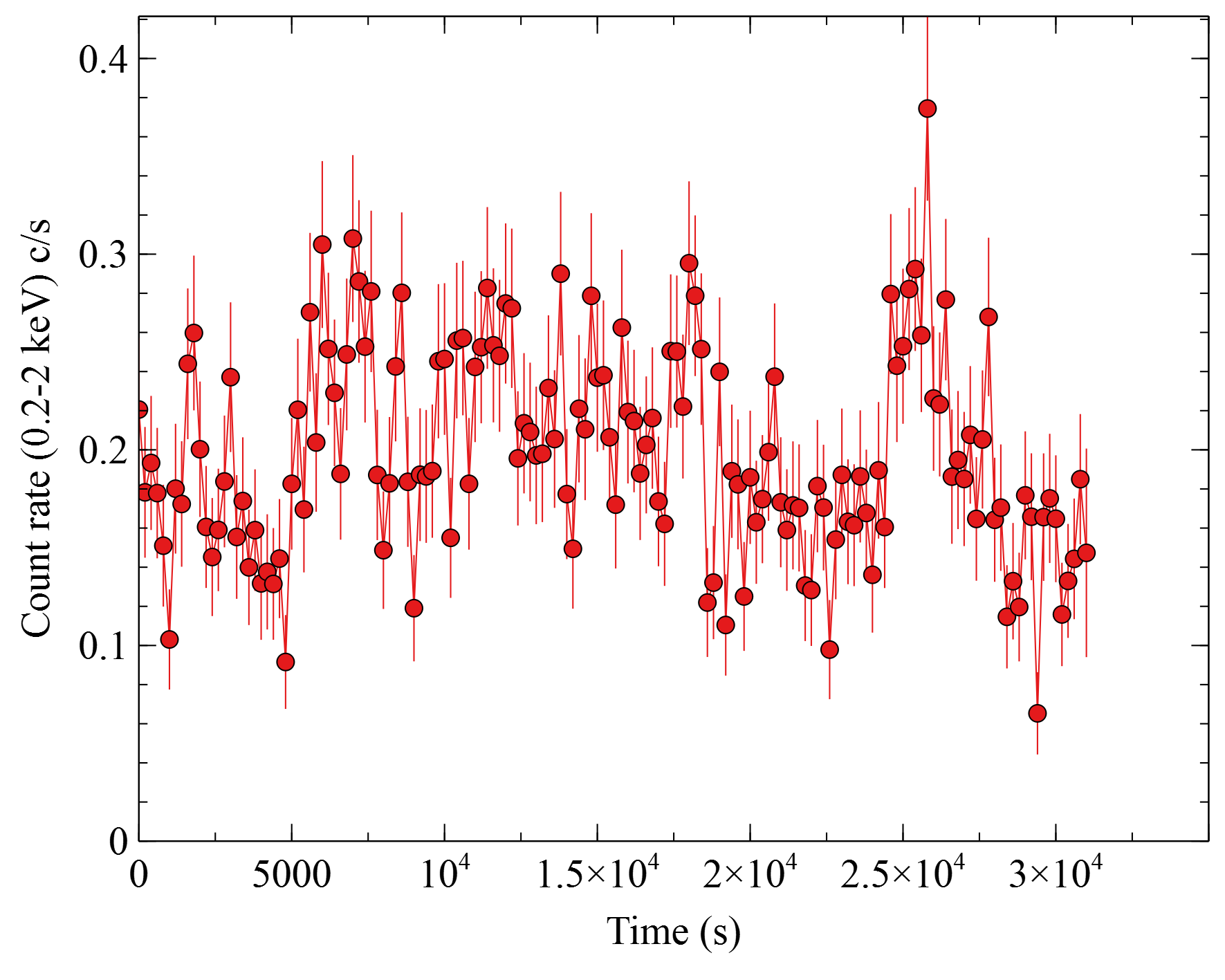}}
    \caption{Soft X-ray light curve of \xsrcname from the \xmm pointed observation of 2018-07-17, binned
    into 200s bins.}
    \label{fig:Xlc_short}
\end{figure}

In Fig.~\ref{fig:Xlc_short} we show the 0.2-2 keV light curve of the 2018-07-17 \xmm observation.
This has a minimum doubling time of about 200s. Applying the relationship between X-ray variability
and mass derived in \citet{Ponti:2012a}
we obtain a black hole mass of $2.5^{+4.2}_{-1.6}\times10^{6}$ \msolarns. From
the K magnitude and the correlation of \citet{MarconiHunt03} we find a mass of
$4.5^{+4.4}_{-2.3}\times10^{6}$\msolarns. 
The mass obtained from fitting the optical spectrum line widths was  
$M_{BH}=5.1^{+9.0}_{-3.2}\times10^{6}$\msolar (Section~\ref{Sec:optspec}). 
We can also use the X-ray spectral properties to estimate $M_{BH}$.
We fit the thermally-dominated, first \swift-XRT observation (S1) with an accretion disc model 
\citep[{\em TDEDISCSPEC};][]{Mummery23}, obtaining a radius $R_{P}=6.1^{+3.3}_{-1.5}\times10^{11}$ cm and peak temperature 
$T_{P}=6.5^{1.0}_{0.6}\times10^{5}$ K ($56^{+9}_{-5}$ eV). From the relationship between 
$R_{P}$ and $M_{BH}$ given in \citet{Mummery23} we derive 
$M_{BH}=3.0^{+4.3}_{-1.8}\times10^{6}$ \msolarns. Finally, using the galaxy stellar mass
of log$M_{*}=10.4$, obtained from the SED fit shown in Fig.~\ref{fig:magellan}, and the relationship 
given in \citep{Greene20}, we derive $M_{BH}=2\times10^{7}$ \msolar with a systematic error of 0.8 dex.

As all mass measurements are consistent within the errors we  
adopt $M_{BH}=4^{+2}_{-2}\times10^{6}$\msolarns.

We barycentre-corrected the light curve, binned it into 10 second bins and
exposure corrected with the {\em epiclccorr} task, from the SAS software \citep{Gabriel04}, to search for any periodicity. None was found.

\renewcommand{\arraystretch}{1.0}

%%%%%%%%%%%%%%%%%%%%%%%%%%%%%%%%%%%%%%%%%%%%%%%%%%

% Don't change these lines
%\bsp  % typesetting comment
\label{lastpage}
\end{document}